\documentclass[graybox]{svmult}

\usepackage{amsmath,amssymb,graphicx,mathtools}
\usepackage{booktabs, multirow}
\usepackage{algorithm,algorithmic,multicol,url,color}

\begin{document}

\title*{Receiving RISs: Enabling Channel Estimation and Autonomous Configuration}
\author{George C. Alexandropoulos\orcidID{0000-0002-6587-1371},\\ Konstantinos D. Katsanos\orcidID{0000-0003-1894-5216}, 
and\\Evangelos Vlachos\orcidID{0000-0003-1501-0722}}

\institute{George C. Alexandropoulos \at Department of Informatics and Telecommunications, National and Kapodistrian University of Athens, Panepistimiopolis Ilissia, 16122 Athens, Greece \email{alexandg@di.uoa.gr}
\and Konstantinos D. Katsanos \at Department of Informatics and Telecommunications, National and Kapodistrian University of Athens, Panepistimiopolis Ilissia, 16122 Athens, Greece \email{kkatsan@di.uoa.gr}
\and Evangelos Vlachos \at Industrial Systems Institute, ATHENA Research and
Innovation Centre, 26504 Rio-Patras, Greece, \email{evlachos@athenarc.gr}}
\maketitle

\abstract{This chapter focuses on a hardware architecture for semi-passive Reconfigurable Intelligent Surfaces (RISs) and investigates its consideration for boosting the performance of Multiple-Input Multiple-Output (MIMO) communication systems. The architecture incorporates a single or multiple radio-frequency chains to receive pilot signals via tunable absorption
phase profiles realized by the metasurface front end, as well as a controller
encompassing a baseband processing unit to carry out channel estimation, and
consequently, the optimization of the RIS reflection coefficients. A novel channel estimation protocol, according to which the RIS receives non-orthogonal training pilot sequences from two multi-antenna terminals via tunable absorption phase profiles, and then, estimates the respective channels via its signal processing unit, is presented. The channel estimates are particularly used by the RIS controller to design the capacity-achieving reflection phase configuration of the metasurface front end. The proposed channel estimation algorithm, which is based on the Alternating Direction Method of Multipliers (ADMM), profits from the RIS random spatial absorption sampling to capture the entire signal space, and exploits the beamspace sparsity and low-rank properties of extremely large MIMO channels, which is particularly relevant for communication systems at the FR3 band and above. Our extensive numerical investigations showcase the superiority of the proposed channel estimation technique over benchmark schemes for various system and RIS hardware configuration parameters, as well as the effectiveness of using
channel estimates at the RIS side to dynamically optimize the possibly phase-quantized reflection coefficients of its unit elements.
}

\section{Introduction}
Reconfigurable Intelligent Surfaces (RISs) constitute an innovative technology of substantial late research and development interest~\cite{EURASIP_RIS_all,BAL_VTM_2024,MML_EMconf_2024} with increased potential to greatly enhance the efficiency of future wireless communication networks~\cite{George_RIS_TWC2019, Marco_Visionary_2019}. Such surfaces, which provide wireless network designers with additional degrees of freedom to intelligently impact signal propagation on the fly, are mainly based on metamaterials and are comprised of periodically aligned subwavelength meta-elements~\cite{ASH_COMMAG_2021}, termed as unit cells or meta-atoms of just elements, which are capable of offering overall control over the metasurface's electromagnetic behavior, ranging from perfect and controllable absorption, that can be used for various purposes~\cite{GGM_ACCESS_2024}, beam and wavefront shaping to polarization control and harmonic generation~\cite{Liaskos_Visionary_2018}.

To effectively operate an RIS for wireless communications, knowledge about the wireless channels between the metasurface and
the communication ends, or the so-called RIS-parametrized cascaded channel, is needed~\cite{Tsinghua_RIS_Tutorial,9771077}. Algorithmic approaches with relevant protocols to estimate the cascaded channel among the base station, RIS, and User Equipment (UE) at either the base station or the UE sides were presented in~\cite{Deepak_RIS_ICASSP2019, He_RIS_CE_2019, Zhang_RIS_CE_2019,9130088,9366805}. However, it was shown that, channel knowledge in RIS-assisted wireless communication setups is, in principle, hard to acquire when the deployed RIS is solely reflective, requiring the need for large overhead channel estimation realized at the end receiver node. On top of this procedure, the channel estimates, as well as the optimized RIS phase profiles based on this estimation, need to be shared with the RIS controller for the latter to optimize\footnote{In certain implementations, the RIS phase profile optimization takes place at the node mastering the RIS, thus, the channel estimates need to be available therein. When the optimization takes place, this node shares the optimized RIS phase profile configuration with the RIS controller.} or set the phase configuration of the metasurface~\cite{SCR_OJCOM_2024}. To reduce the channel estimation overhead, techniques that exploit the channel structure and/or deploy RIS pilot phase profiles have also been proposed~\cite{LHA_TCOM_2021,LZA_TWC_2021,LHG_TWC_2022,9732214,BGA_TWC_2024}. For example, in~\cite{9732214}, a low pilot overhead technique was presented exploiting the known positions of the UEs and the base station. Last but not least, RIS configuration optimization schemes that avoid explicit estimation of the involved channel matrices have been also designed. In particular, phase profile codebooks and phase configuration training protocols were presented in~\cite{YBZ_WCL_2020,AJS_SAM_2022,WFZ_WCOM_2022,RDK_EURASIP_2023}, while~\cite{JAR_COML_2022} optimized the RIS phase response according to an area illumination criterion only when necessary (e.g., when communication performance drops a certain quality-of-service threshold), and performed conventional estimation of the RIS-parameterized Multiple-Input Multiple-Output (MIMO) channel on a per channel access basis. Differential data-aided RIS phase profile training and non-coherent modulation with random RIS phase configurations were designed in~\cite{CAG_ACCESS_2022} and~\cite{CAG_ITU_2022}, respectively.

To enable RISs perform estimation of parameters of their
impinging waveforms, thus, facilitating and expediting their optimization for wireless communications and sensing, an RIS hardware architecture incorporating Reception (RX) Radio-Frequency (RF) chains (also termed as receiving RIS (emphasizing its reception capability) or semi-passive RIS (emphasizing the inclusion of active reception components)), which are fed with the impinging signals on the RIS unit elements when those are configured on a full absorption mode, was proposed in~\cite{AV_ICASSP_2020}. This architecture that also includes a baseband processor was deployed to perform channel estimation at the RIS side via a low pilot overhead scheme that exploits the sparsity of large MIMO channels at millimeter wave, and beyond, frequencies, as well as for signal direction estimation~\cite{locrxris_all}. Another semi-passive RIS hardware architecture comprising conventional solely reflective unit elements and active sensors, that are connected to baseband feeding a channel estimation mechanism, was presented in~\cite{Alkhateeb_RIS_CS_2019}. Very recently, in~\cite{ASI_VTM_2024}, an RIS hardware architecture comprising hybrid unit elements capable of simultaneous tunable reflection and tunable absorption was presented that has been also used for efficiently estimating the cascaded channel in RIS-assisted multi-user communication systems~\cite{ZSA_TCOM_2023}.

\subsection{Chapter's Contribution}
In this chapter, we capitalize on~\cite{AV_ICASSP_2020}'s RIS hardware architecture and study its consideration for boosting the performance of MIMO communication systems. In particular, we consider a system model comprising a receiving RIS capable of channel estimation and two multi-antenna UEs, which wish to profit from RIS-enabled tunable reflections to enhance the performance of their end-to-end MIMO communication link. We assume that the RIS incorporates multiple RX RF chains to receive pilot signals via tunable absorption phase profiles realized by the metasurface, as well as that the RIS controller encompasses a baseband processing unit to carry out channel estimation, and consequently, the optimization of the RIS reflection coefficients. Our specific contributions can be summarized as follows.
\begin{itemize}
    \item We present a novel system model including two multi-antenna UEs that interact with a receiving RIS placed in their vicinity through a dedicated control channel, expressing their willingness to profit from RIS reflection optimization. We design a channel estimation phase according to which the metasurface receives non-orthogonal training pilot sequences via tunable absorption phase profiles, and then, estimates the respective channels via its processing unit. The channel estimates are then used by the RIS controller to design the capacity-achieving reflection coefficients of the metasurface.
    \item The considered RIS hardware architecture uses random spatial absorption sampling to capture the entire signal space, without requiring knowledge of the position of either of the UEs. This feature reduces the initialization phase for joint channel estimation of multiple UEs compared to orthogonal channel estimation schemes. 
    \item We design a novel channel estimation approach exploiting the beamspace sparsity and low-rank properties of large MIMO channels, which is particularly relevant for high-frequency communication systems. This approach extends the techniques proposed in~\cite{Vlachos_SPL2018,Vlachos2019WidebandSampling} to partially-connected receiving RISs and dynamic metasurface antennas~\cite{SAI_WCOM_2021}, thus, being more efficient. Our solution is based on the Alternating Direction Method of Multipliers (ADMM) and is optimized with fewer steps than the state of the art, thus, being computationally more efficient.
    \item Based on the individual channel estimates available at the RIS controller, we present a capacity-achieving RIS phase configuration scheme for the end-to-end communication between the two UEs, which is assumed to be entirely implemented at the RIS controller.
    \item We present an extensive performance evaluation of the proposed channel estimation technique, including a convergence analysis, as well as of the devised self reflection configuration scheme for the considered semi-passive RIS architecture, showcasing the interplay of various system, channel, and channel estimation parameters on the overall performance.
\end{itemize}

\subsection{Chapter Organization and Notations}
The chapter is organized as follows. In Section~\ref{sec:System_Channel_Models}, the system and channel models are introduced. In particular, the key RIS hardware components and mode of operation as well as the protocol for channel estimation and autonomous phase configuration at the RIS side are first presented, which are followed by the received signal model at the semi-passive RIS as well as the RIS reflection phase configuration optimization objective. Next, the matrix mathematical models for both MIMO channels between each of the UEs and the RIS are presented. In Section~\ref{sec:Channel_Estimation}, the proposed channel estimation problem is described that capitalizes on the receiving RIS architecture which is tasked to realize random spatial absorption sampling to capture the entire signal space. The RIS-side channel estimation problem is specifically formulated as a joint low rank-sparsity optimization problem. Then, via the ADMM method, the problem is broken into two separate sub-problems. Mathematical application of the optimality conditions leads to one of those sub-problems to take the form of a matrix completion problem, which is efficiently solved via the Singular Value Thresholding (SVT) method. The other sub-problem takes the form of a Least Absolute Shrinkage and Selection Operator (LASSO) problem~\cite{Tibshirani1996} which is solved via the soft-thresholding operator. The convergence analysis of the overall channel estimation algorithm is also included, following the convergence properties of SVT and LASSO. In Section~\ref{sec:Results}, the performance of the proposed channel estimation algorithm is numerically evaluated, together with the proposed autonomous RIS reflection phase configuration optimization approach. In particular, results on the convergence rate of the designed ADMM-based channel estimation and its Mean Square Error (NMSE) performance as a function of the Signal-to-Noise Ration (SNR), the number of training symbols, and the number of the RX RF chains at the considered semi-passive RIS, are presented. Performance comparisons with benchmark channel estimation schemes are also included, and the impact of channel estimation on the capacity-achieving RIS reflection phase configuration is numerically investigated. Finally, Section~\ref{sec:conclusions} lists the concluding remarks of the chapter together with relevant directions for future research.

A summary of the notation used throughout this chapter is provided in Table~\ref{table:notation}.
\begin{table}[!t]
  \centering
  \caption{The Notations of this Chapter.}
  \begin{tabular}{r|l}
    \toprule
    $a, \mathbf{a}$, and $\mathbf{A}$ &  Scalar, vector, and matrix \\
    $\jmath \triangleq \sqrt{-1}$ & The imaginary unit \\
    $\texttt{Re}(\cdot)$ & The real part of a complex scalar/vector/matrix\\
    $\mathtt{Im}(\mathbf{\cdot})$ & The imaginary part of a complex scalar/vector/matrix \\
    $\mathbf{A}^{\rm T}$ and $\mathbf{A}^{\rm H}$ &  Matrix transpose and Hermitian transpose \\
    $\mathbf{A}^{-1}$ and $\mathbf{A}^{\dagger}$ &  Matrix inverse and pseudo-inverse \\    
    $[\mathbf{A}]_{i,j}$ &  Matrix element at the $i$-th row and $j$-th column \\
    $[\mathbf{a}]_{i}$ &  The $i$-th vector element \\
    $\mathbf{A}$ and $\mathbf{\hat{A}}$ & Actual and estimated matrix \\
    $\mathbf{I}_N$ & $N \times N$ identity matrix \\
    $\mathbf{0}_{N \times K}$ & $N \times K$ matrix with zeros\\
    $\mathbf{I}_{N \times K}$ & Column concatenated matrix $[\mathbf{I}_N \,\, \mathbf{0}_{N \times K}]$ \\
    $\mathbf{\Omega}$ & Matrix containing $0$'s and $1$'s \\
    $\Vert \cdot \Vert_{\rm F}$ & Matrix Frobenius norm \\
        $\Vert \cdot \Vert_*$ & Matrix Nuclear norm \\
                $\Vert \mathbf{a} \Vert_m$ & $l_{m}$ norm of vector $\mathbf{a}$\\
    $\times$ & Scalar multiplication \\
    $\circ$ & Element-wise (Hadamard) matrix product \\
    $\otimes$ & Kronecker product \\
    $\textrm{tr}(\mathbf{A})$ & The trace of matrix $\mathbf{A}$ \\
    $\textrm{vec}(\mathbf{A})$ & Vectorization of $\mathbf{A}$ \\    
    $\textrm{vec}_{\rm d}(\mathbf{A})$ & Vector whose elements are the diagonal entries of the square matrix $\mathbf{A}$ \\
    $\textrm{unvec}(\cdot)$ & Inverse operation of $\textrm{vec}(\cdot)$ \\    
    $\textrm{diag}(\mathbf{a})$ & Diagonal matrix with $\mathbf{a}$ on the main diagonal\\
    $\nabla_{\mathbf{a}} f(\mathbf{a})$ & Gradient vector of function $f(\mathbf{a})$ with respect to $\mathbf{a}$ \\
    $\text{sgn}\{\mathbf{a}\}$ & Sign of $\mathbf{a}$\\
    \bottomrule
  \end{tabular}
  \label{table:notation}
\end{table}

\section{System and Channel Models}\label{sec:System_Channel_Models}
In this section, we describe the RIS-assisted MIMO communication system model and the channel model considered in this chapter. The system model comprises an RIS and two multi-antenna UEs, where the latter wish to profit from RIS-enabled optimizable generalized reflection configurations to enhance the characteristics of their communication link. We consider that the RIS is implemented via the hardware architecture of~\cite{SAI_WCOM_2021}, which includes RX RF chains connecting all dynamically controlled front-end metamaterials with a baseband processing unit incorporated to the RIS controller. This architecture enables the overall metasurface to receive and process pilot-carrying signals from both the multi-antenna UEs. To this end, in a first phase, both of them communicate to the RIS controller, through a dedicated control channel~\cite{EURASIP_RIS_all,FCR_OJCOM_2024}, their willingness to use the metasurface for enhancing their communication performance, and transmit pilot symbols in a synchronized manner. The received pilot signals at the semi-passive RIS side are used by its baseband processing unit to estimate the respective channels. Those estimations are then used by this unit to first formulate the communication objective and then compute accordingly the optimized RIS reflection coefficients for the inter-UE communication~\cite{ASI_VTM_2024}. Specifically, this optimized reflection phase configuration is set at a second phase where data communications between the UEs take place. 

\subsection{System Model and RIS Mode of Operation}\label{sec:Model_Mode_or_Operation}
\begin{figure}[!t]
	\centering
	\includegraphics[width=0.8\columnwidth]{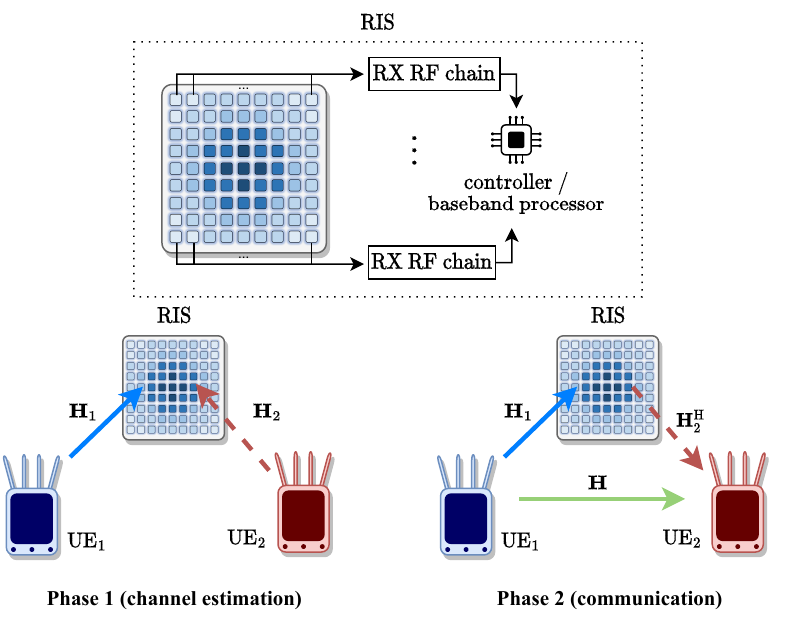} 
	  \caption{
   The considered communication system model including an $N_{\rm RIS}$-element RIS capable to receive pilot signals via its $N_{\rm RF}\ll N_{\rm RIS}$ Reception (RX) Radio-Frequency (RF) chains, perform channel estimation through its baseband processor, and consequently, optimize its reflection phase configuration to assist the wireless link between the $N_1$-antenna node UE$_1$ and the $N_2$-antenna node UE$_2$. The system operates in a Time Division Duplexing (TDD) manner including two phases: one phase for the estimation of the UE$_1$-to-RIS and UE$_2$-to-RIS channels $\mathbf{H}_1$ and $\mathbf{H}_2$, respectively, which is followed by another phase dedicated for inter-UE data communications assisted by RIS-optimized reflection. The detailed communication protocol is sketched in Fig.~\ref{fig:R-RIS_protocol}.}
		\label{fig:R-RIS_system_model}
\end{figure}
The considered system model is illustrated in Fig.~\ref{fig:R-RIS_system_model} consisting of the multi-antenna UE$_1$ ($N_1$ elements) and UE$_2$ ($N_2$ elements) wishing to establish wireless communications through the optimization of a semi-passive RIS installed in their vicinity, whose position is such that it can impact the link's quality~\cite[Sec.~IV.B]{MAD_TWC_2023}. It is assumed that the RIS can realize both tunable absorption and tunable reflection in discrete time instances, with the former operation being performed so as to guide the impinging signals upon the metasurface to a baseband processor embedded in the RIS controller. The RIS is implemented similar to the hardware architecture of~\cite{SAI_WCOM_2021}, in particular, it consists of $N_{\rm RIS}$ sub-wavelength-spaced metamaterials of tunable responses which are coated in $N_{\rm E}$-element disjointed groups to $N_{\rm RF}$ distinct waveguides, with each of the latter attached to an RX RF chain (typically composed of a low noise amplifier,
a mixer which downconverts the signal from RF to baseband,
and an analog-to-digital converter~\cite{AIS_VTM_2022}). The outputs of the $N_{\rm RF}\ll N_{\rm RIS}(=N_{\rm RF}N_{\rm E})$ RX RF chains feed the receiving RIS's baseband processor, which is tasked to perform channel estimation and phase configuration optimization. 

\begin{figure}[!t]
	\centering
	\includegraphics[width=0.9\columnwidth]{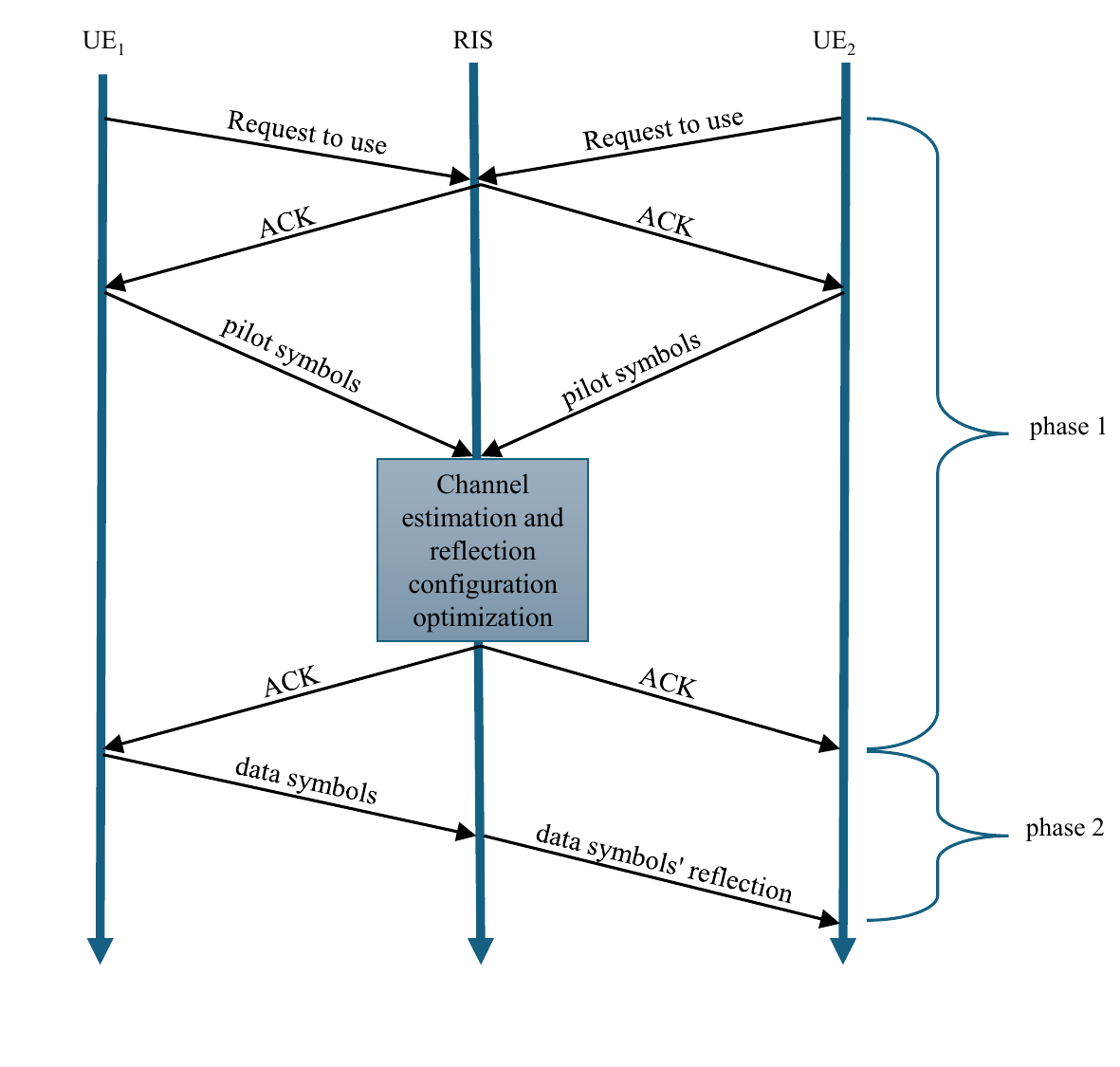} 
	  \caption{
   The proposed communication protocol between the multi-antenna UE$_1$ and UE$_2$ incorporating the considered semi-passive RIS device, which is capable of pilot symbols' reception via tunable signal absorption configuration, channel estimation, and tunable signal reflection configuration. The RIS device comprises the metasurface panel and the RIS controller that includes a baseband processing unit~\cite{RIS_AIoT_2025}.}
		\label{fig:R-RIS_protocol}
\end{figure}
Our RIS-empowered MIMO communication system model operates similar to~\cite{AV_ICASSP_2020} in a Time Division Duplexing (TDD) manner. In the first phase, the nodes UE$_1$ and UE$_2$ that wish to profit from the RIS reflection phase optimization inform, in a dedicated control channel~\cite{EURASIP_RIS_all,FCR_OJCOM_2024}, the RIS controller about their intention to engage the metasurface for enabling/boosting their data communication. The latter follows back by sending an ACKnowledgement (ACK) notifying its readiness to contribute to their communication link, and both UE$_1$ and UE$_2$ begin transmitting training pilot symbols simultaneously in a synchronized manner. Upon reception of the pilot-bearing signals at the RIS baseband processing unit, enabled by the considered semi-passive RIS architecture incorporating tunable absorption phase configuration and RX RF chains, channel estimation is performed which is then used for the optimization of the RIS reflection phase configuration. When the latter process is finalized, the RIS device transmits an ACK to both UE$_1$ and UE$_2$ to kickstart the data communication phase. This communication begins with the RIS realizing optimized generalized reflection of its information-bearing impinging signals. This two-phase communication protocol between UE$_1$ and UE$_2$ profiting from the considered semi-passive RIS is schematically illustrated in Fig.~\ref{fig:R-RIS_protocol}.   

\subsubsection{Received Signal Model}
Let $2T$ denote the total number of pilot symbols transmitted by UE$_1$ and UE$_2$ in $T$ consecutive discrete time slots within each channel coherence time for both UE$_1$-RIS and UE$_2$-RIS channel estimations. At each $t$-th time slot, with $t=1,2,\ldots,T$, within a coherence block of channels, the complex-valued $N_{\rm RIS}$-element vector including the baseband versions of the signals received at the RIS unit elements can be mathematically expressed as follows:
\begin{align}
    \mathbf{y}(t) 
    = \mathbf{H}_1 \mathbf{s}_1(t) + \mathbf{H}_2 \mathbf{s}_2(t), \label{eq:received_RIS_elements}
\end{align}
where $\mathbf{s}_1(t) \in \mathbb{C}^{N_1 \times 1}$ and $\mathbf{s}_2(t) \in \mathbb{C}^{N_2 \times 1}$ represent respectively the transmitted pilot symbols from UE$_1$ and UE$_2$ during each $t$-th time slot, with $\Vert \mathbf{s}_1(t) \Vert^2_2$ and $\Vert \mathbf{s}_2(t) \Vert^2_2$ being the respective transmit powers. In addition, $\mathbf{H}_1 \in \mathbb{C}^{N_{\rm RIS} \times N_1}$ and $\mathbf{H}_2 \in \mathbb{C}^{N_{\rm RIS} \times N_2}$ include the gains of the UE$_1$-to-RIS and UE$_2$-to-RIS channels, respectively. 

We define the $N_{\rm RIS}\times N_{\rm RIS}$ diagonal matrix $\mathbf{P}$, with elements modeling signal propagation inside the waveguides hosting the $N_{\rm RF}$ groups of $N_{\rm E}$ metamaterials, as follows $\forall$$i=1,2,\dots,N_{\rm RF}$ and $\forall$$n = 1,2,\dots,N_{\rm E}$~\cite{XYA_TWC_2024}:
\begin{align}\label{eq: TX_Sig_Prop}
    [\mathbf{P}]_{(i-1)N_{\rm E}+n,(i-1)N_{\rm E}+n} \triangleq e^{-\rho_{i,n}(\alpha_i + \jmath\beta_i)},
\end{align}
where $\alpha_i$ represents the waveguide attenuation coefficient, $\beta_i$ stands for the wavenumber, and $\rho_{i,n}$ signifies the position of the $n$-th element within the $i$-th waveguide\footnote{This model can be easily extended to incorporate losses within each waveguide~\cite{DMA_Losses_2025}. The framework presented in this chapter holds for any of the available models for the $\mathbf{P}$ matrix.}. Note that this matrix depends on the physical characteristics of the waveguides and those of the RIS unit elements. Let also $u_{i,n}(t)$ be the adjustable absorption phase response (i.e., analog combining weight), associated with each $n$-th metamaterial in each $i$-th waveguide during each $t$-th time slot. This weight is assumed to conform to a Lorentzian-constrained phase model and is assumed to belong to the phase profile codebook $\mathcal{W}_{\rm a}$, i.e.:
\begin{align}\label{eq: code}
    u_{i,n}(t) \in \mathcal{W}_{\rm a}\triangleq \left\{0.5\left(\jmath+e^{\jmath\varphi_{\rm a}}\right)\Big|\varphi_{\rm a}\in\left\{2^{2-b}\pi m\right\}_{m=0}^{2^{b-1}}\right\}
\end{align}
including in total $|\mathcal{W}_{\rm a}|=2^{b-1}+1$ absorption phase states. Using this definition, the matrix $\mathbf{U}(t)\in\mathbb{C}^{N_{\rm RIS}\times N_{\rm RF}}$ with the tunable absorption coefficients of the RIS unit elements (i.e., the RIS partially-connected analog combining matrix for the impinging pilot signals) at each $t$-th time slot can be expressed as follows ($\forall$$j=1,2,\dots,N_{\rm RF}$):
\begin{align}\label{eq:Us}
    [\mathbf{U}(t)]_{(i-1)N_{\rm E}+n,j} = \begin{cases}
    u_{i,n}(t),&  i=j\\
    0,              & i\neq j
\end{cases}.
\end{align}

Using the latter definitions as well as expression~\eqref{eq:received_RIS_elements} for the impinging pilot signals on the metasurface, the received complex-valued $N_{\rm RF}$-element vector at each $t$-th time slot available at the RIS baseband processor, to be processed for channel parameter estimation purposes, is given by:
\begin{align} 
     \mathbf{y}_{\rm RF}(t)  
    &= \mathbf{U}^{\rm T}(t) \underbrace{\mathbf{P}^{\rm H} \left(\mathbf{H}_1 \mathbf{s}_1(t) + \mathbf{H}_2 \mathbf{s}_2(t)\right)}_{\triangleq \mathbf{y}_{\rm RIS}(t)}+\mathbf{n}_{\rm RF}(t),
    \label{eq:received_RIS_RF_chains}
\end{align}
where $\mathbf{n}_{\rm RF}(t) \in \mathbb{C}^{N_{\rm RF} \times 1}$ is the zero-mean complex Additive White Gaussian Noise (AWGN) vector for the $t$-th pilot symbol with covariance matrix $\sigma^2\mathbf{I}_{N_{\rm RF}}$, i.e., $\mathbf{n}_{\rm RF}(t) \sim \mathcal{CN}(\mathbf{0}_{N_{\rm RF}}, \sigma^2 \mathbf{I}_{N_{\rm RF}})$. Notation $\mathbf{y}_{\rm RIS}(t)\in \mathbb{C}^{N_{\rm RIS} \times 1}$ in the expression~\eqref{eq:received_RIS_RF_chains} includes the signals impinging at the RIS unit elements at this $t$-th time interval.

\subsubsection{RIS Reflection Phase Optimization}\label{sec:RIS_reflection_optimization}
After computing the estimates 
for the channels $\mathbf{H}_1$ and $\mathbf{H}_2$, 
the RIS baseband processor solves the following capacity optimization problem to obtain the optimum RIS reflection phase configuration for assisting UE$_1$-to-UE$_2$ communications; we focus on the case where UE$_1$ intends to transmit power-limited Gaussian-distributed data symbols 
to UE$_2$ (reversing the roles of UE$_1$ (transmitter) and UE$_2$ (receiver) will lead to a similar optimization problem):
\begin{align*}
\mathcal{OP}_1\!: \,&\underset{\boldsymbol{\varphi}}{\max} \log_2\left(\det\left(\mathbf{I}_{N_2}+{\rm SNR}\mathbf{H}_{\rm e2e}(\boldsymbol{\varphi})\mathbf{H}_{\rm e2e}^{\rm H}(\boldsymbol{\varphi})\right)\right)\\&{\rm s.t.}\,\,[\boldsymbol{\varphi}]_i\in\mathcal{W}_{\rm r}\,\,\forall i=1,2,\ldots,N_{\rm RIS},
\end{align*}
where ${\rm SNR}$ is the ratio between the transmit power and the AWGN variance, and the $N_{\rm RIS}$-element vector $\boldsymbol{\varphi}$ includes the RIS tunable reflection coefficients, i.e., $\forall i$:
\begin{equation}
[\boldsymbol{\varphi}]_i\in\mathcal{W}_{\rm r}\triangleq \left\{0.5\left(\jmath+e^{\jmath\varphi_{\rm r}}\right)\Big|\varphi_{\rm r}\in\left\{2^{2-b}\pi m\right\}_{m=0}^{2^{b-1}}\right\} 
\end{equation}
including in total\footnote{Without loss of generality, the reflection phase profile codebook $\mathcal{W}_{\rm r}$ has been considered to have the same elements and cardinality with the absorption one, $\mathcal{W}_{\rm a}$, in~\eqref{eq: code}. In general, these codebooks can be different (both in the actual elements and their total numbers) due to different constraints and requirements for the respective functionalities.} $|\mathcal{W}_{\rm r}|=2^{b-1}+1$ reflection states. In addition, $\mathbf{H}_{\rm e2e}(\boldsymbol{\varphi})\triangleq \mathbf{H}+\mathbf{H}_2^{\rm H}{\rm diag}(\boldsymbol{\varphi})\mathbf{H}_1\in \mathbb{C}^{N_2 \times N_1}$ in $\mathcal{OP}_1$ represents the overall RIS-parameterized UE$_2$-to-UE$_1$ channel matrix for the data communication phase. In the latter expression, $\mathbf{H}\in\mathbb{C}^{N_2 \times N_1}$ denotes the direct UE$_2$-to-UE$_1$ channel\footnote{This channel can be estimated at the receiver of UE$_2$ via conventional MIMO channel estimation techniques~\cite{heath2018foundations}.} and $\mathbf{H}_2^{\rm H}$ indicates $\mathbf{H}_2$'s reciprocal version, i.e., from the RIS to UE$_2$. In $\mathcal{OP}_1$, we have assumed, without loss of generality, that the noise variance related to each UE$_2$ RX antenna element is the same.

The authors in~\cite{zhang2020capacity} considered the RIS-aided point-to-point MIMO system and studied the capacity limits for both narrowband and generic broadband Orthogonal Frequency Division Multiplexing (OFDM) cases, taking into account continuous values for the RIS reflection coefficients. The same system was also investigated in~\cite{perovic2021achievable}, where the focus was on the design of the transmit covariance matrix along with the RIS reflection coefficients. Therein, a method based on the projected gradient was designed that required lower computational complexity. The MIMO capacity maximization problem was also studied in~\cite{sirojuddin2024mimo}, considering a more practical model for the RIS phase response that was based on the equivalent circuit model for each unit element. The solutions for the optimal power allocation and RIS reflection coefficients were obtained using an iterative algorithm based on the Newton-Raphson method and a modified gradient ascent for each subproblem. The conducted numerical investigations demonstrated that the proposed method outperforms benchmark schemes in terms of achievable rate and execution time. On the other hand,~\cite{AV_ICASSP_2020,HAZ2018,wu2019beamforming,di2020hybrid} studied the case of optimizing an RIS which admits discrete states for its reflection coefficients, as we consider in $\mathcal{OP}_1$. In particular,~\cite{AV_ICASSP_2020} solved a special case of $\mathcal{OP}_1$ where UE$_1$ and UE$_2$ were equipped with a single antenna each. The authors in~\cite{wu2019beamforming} studied the transmit power minimization problem with respect to the precoding at the transmitter and the discrete RIS reflection phase profiles, enforcing constraints on the minimum signal-to-noise-plus-interference ratios at the receivers, leading to a mixed-integer non-linear problem. On a similar basis, a hybrid beamforming scheme for the sum-rate maximization problem in a downlink multi-user system, where the transmission is aided by an RIS was investigated in~\cite{di2020hybrid}, showcasing that the proposed design therein can achieve good performance even for a low-resolution set of values for the discrete RISs. 

The $\mathcal{OP}_1$ formulation is an intractable problem, because the objective function is non-convex with respect to $\boldsymbol{\varphi}$ and the constraints admit discrete values. However, it can be iteratively solved based on a gradient ascent scheme by relaxing the constraint set to the continuous one $\mathcal{W}_{\rm r,c} \triangleq \{0.5(\jmath + e^{\jmath\bar{\varphi}})|\bar{\varphi}\in[0,2\pi]\}$. Then, a suboptimal solution can be obtained by performing quantization, such that it belongs to the set $\mathcal{W}_{\rm r}$. To proceed to the solution, we first need to obtain the gradient of the objective function with respect to the $N_{\rm RIS}$-element vector $\bar{\boldsymbol{\varphi}}$ with $[\bar{\boldsymbol{\varphi}}]_i\in\mathcal{W}_{{\rm r,c}}$ $\forall$$i$. To this end, let $f(\bar{\boldsymbol{\varphi}}) \triangleq \log_2\left(\det\left(\mathbf{I}_{N_2}+{\rm SNR}\mathbf{H}_{\rm e2e}(\bar{\boldsymbol{\varphi}})\mathbf{H}_{\rm e2e}^{\rm H}(\bar{\boldsymbol{\varphi}})\right)\right)$. Then, it can be shown that, for the derivative of this function with respect to $\bar{\boldsymbol{\varphi}}$, holds:
\begin{equation}
\nabla_{\bar{\boldsymbol{\varphi}}}f(\bar{\boldsymbol{\varphi}}) = 2\frac{{\rm SNR}}{\ln(2)}\texttt{Re}\left(\operatorname{diag}(\mathbf{v})\operatorname{vec}_{\rm d}\left(\mathbf{H}_1 \mathbf{H_{\rm e2e}^{\rm H}}(\bar{\boldsymbol{\varphi}})\mathbf{R}^{-1}(\bar{\boldsymbol{\varphi}}) \mathbf{H}_2^{\rm H}\right)\right)
\end{equation}
with $\mathbf{v} \triangleq \jmath 0.5\left[e^{\jmath \bar{\varphi}_1},e^{\jmath \bar{\varphi}_2},\ldots,e^{\jmath \bar{\varphi}_{N_{\rm RIS}}}\right]^{\rm T}$ and $\mathbf{R}(\bar{\boldsymbol{\varphi}}) \triangleq \mathbf{I}_{N_2}+{\rm SNR}\mathbf{H}_{\rm e2e}(\bar{\boldsymbol{\varphi}})\mathbf{H}_{\rm e2e}^{\rm H}(\bar{\boldsymbol{\varphi}})$. Each $n$-th iteration of the projected gradient ascent algorithm, with $\mu>0$ being the step size, is given by:
\begin{equation}\label{eq:iteration}
	\bar{\boldsymbol{\varphi}}^{(n+1)} = \left(\bar{\boldsymbol{\varphi}}^{(n)} + \mu \nabla_{\bar{\boldsymbol{\varphi}}}f(\bar{\boldsymbol{\varphi}})|_{\bar{\boldsymbol{\varphi}}=\bar{\boldsymbol{\varphi}}^{(n)}}\right)\mod{2\pi}.
\end{equation}
Finally, as previously described, it suffices to solve the following optimization problem to obtain the optimized for data communication $[\boldsymbol{\varphi}]_i\in \mathcal{W}_{\rm r}$ $\forall$$i$:
\begin{equation}
	[\boldsymbol{\varphi}]_i =  \underset{\kappa\in\mathcal{W}_{\rm r}}{\arg\min} \left|\left[\bar{\boldsymbol{\varphi}}^{(I_{\max})}\right]_i - \kappa\right|,
\end{equation}
where $I_{\max}$ indicates the iteration index where \eqref{eq:iteration} converges, or the maximum number of iterations. 

The overall design for the RIS reflection phase configuration $\bar{\boldsymbol{\varphi}}$ solving $\mathcal{OP}_1$ is summarized in Algorithm~\ref{algorithm:RIS_optimization}. 
\begin{algorithm}[!t]
	\caption{Design of the RIS Reflection Phase Configuration $\bar{\boldsymbol{\varphi}}$}
	\begin{algorithmic}[1]
		\REQUIRE ${\rm SNR}$, $\mathbf{H}$, $\mathbf{H}_1$, $\mathbf{H}_2$, $I_{\rm max}$, arbitrary $\bar{\boldsymbol{\varphi}}^{(0)}\in[0,2\pi]$.
		\ENSURE $\bar{\boldsymbol{\varphi}}^{(I_{\rm max}+1)}$.
		\FOR {$n=0,1,\ldots,I_{\rm max}$}     

        \STATE Compute $\mathbf{R}\left(\bar{\boldsymbol{\varphi}}^{(n)}\right) = \mathbf{I}_{N_2}+{\rm SNR}\mathbf{H}_{\rm e2e}\left(\bar{\boldsymbol{\varphi}}^{(n)}\right)\mathbf{H}_{\rm e2e}^{\rm H}\left(\bar{\boldsymbol{\varphi}}^{(n)}\right)$ 
        
        with $\mathbf{H}_{\rm e2e}\left(\bar{\boldsymbol{\varphi}}^{(n)}\right)= \mathbf{H}+\mathbf{H}_2^{\rm H}{\rm diag}\left(\bar{\boldsymbol{\varphi}}^{(n)}\right)\mathbf{H}_1$. \\
        \STATE Set $\mathbf{v} = \jmath 0.5\left[e^{\jmath \bar{\varphi}_1^{(n)}},e^{\jmath \bar{\varphi}_2^{(n)}},\ldots,e^{\jmath \bar{\varphi}_{N_{\rm RIS}}^{(n)}}\right]^{\rm T}$. \\
        \STATE Compute $\nabla_{\boldsymbol{x}}f(\boldsymbol{x}) = 2\frac{{\rm SNR}}{\ln(2)}\texttt{Re}\left(\operatorname{diag}(\mathbf{v})\operatorname{vec}_{\rm d}\left(\mathbf{H}_1 \mathbf{H_{\rm e2e}^{\rm H}}(\boldsymbol{x})\mathbf{R}^{-1}(\boldsymbol{x}) \mathbf{H}_2^{\rm H}\right)\right)$ \\
        \STATE Set $\bar{\boldsymbol{\varphi}}^{(n+1)} = \left(\bar{\boldsymbol{\varphi}}^{(n)} + \mu \nabla_{\boldsymbol{x}}f(\boldsymbol{x})|_{\boldsymbol{x}=\bar{\boldsymbol{\varphi}}^{(n)}}\right)\mod{2\pi}$. \\
        \IF{Convergence criterion is met for the rate function}
        \STATE break;
        \ENDIF
        \ENDFOR
	\end{algorithmic}
	\label{algorithm:RIS_optimization}
\end{algorithm}

\subsection{Channel Model}
Both UE$_1$ and UE$_2$ are equipped with a Uniform Linear Arrays (ULA) consisting of $N_1$ and $N_2$ antenna elements, respectively. On the other hand, the RIS is a Uniform Planar Array (UPA) of $N_{\rm RIS,y}$ tunable absorption/reflection metamaterials in the y-axis and $N_{\rm RIS,z}$ ones in the z-axis, such that $N_{\rm RIS}=N_{\rm RIS,y} \times N_{\rm RIS,z}$. The matrix $\mathbf{H}_1$ with the UE$_1$-RIS instantaneous channel gain coefficients, assuming $P$ propagation paths, is modeled as follows:
\begin{equation}
    \mathbf{H}_1 \triangleq \sqrt{\frac{N_1 N_{\rm RIS}}{P}} \sum_{p=1}^P \alpha_p  \mathbf{a}_{\rm RIS}(\theta_{\rm R}^p, \phi^p) \mathbf{a}_{1}^{\rm H}(\theta_{\rm T}^p),\label{eq1}
\end{equation}
where each $\alpha_p$ denotes the complex-valued gain of the $p$-th UE$_1$-RIS channel component having the azimuth Angle of Arrival (AoA) at the RIS represented by $\theta_{\rm R}^p$, the azimuth Angle of departure (AoD) from the UE$_1$ by $\theta_{\rm T}^p$, and $\phi^p$ indicates the normalized elevation AoA at the RIS. Note that the normalized spatial AoA $\theta \in [-1,1]$ is related to the physical $\tilde{\theta} \in [-\frac{\pi}{2}, \frac{\pi}{2}]$ via the expression $\theta = \frac{d}{\lambda} \sin \tilde{\theta}$, where $d$ is the antenna spacing and $\lambda$ represents the signal wavelength. Similarly, the normalized spatial AoD $\phi$ is related to the physical $\tilde{\phi} \in [-\frac{\pi}{4}, \frac{\pi}{4}]$ via the expression $\phi = \frac{d}{\lambda} \sin \tilde{\phi}$. In addition, the $N_1\times1$ steering vector at the UE$_1$ and the $N_{\rm RIS}\times1$ steering vector at the considered semi-passive RIS are given respectively by:
\begin{align}
    \mathbf{a}_{1}(\theta) &\triangleq \boldsymbol{\beta}(\cos \theta, N_1),\label{eq:steering_UE1}\\
     \mathbf{a}_{\rm RIS}(\theta, \phi) &\triangleq \mathbf{a}_y(\theta,\phi)\otimes \mathbf{a}_z(\phi),
\end{align}
where for $x \in [-1,1]$: 
\begin{equation}
    \boldsymbol{\beta}(x, N) \triangleq \frac{1}{\sqrt{N}} [ 1, e^{-\jmath \pi x},e^{-\jmath 2\pi x}, \ldots, e^{-\jmath \pi (N-1)x}]^{\rm T},
\end{equation}
and $\mathbf{a}_y(\theta,\phi)\triangleq\boldsymbol{\beta}(\sin \theta \sin \phi, N_{\rm RIS, y})$ and $\mathbf{a}_z(\phi)\triangleq\boldsymbol{\beta}(\cos \phi, N_{\rm RIS, z})$.

Similar to $\mathbf{H}_1$, the matrix $\mathbf{H}_2$ with the UE$_2$-RIS instantaneous channel gain coefficients is modeled assuming $Q$ propagation paths as follows:
\begin{equation}
    \mathbf{H}_2 \triangleq \sqrt{\frac{N_{\rm 2} N_{\rm RIS}}{Q}} \sum_{q=1}^Q \beta_{q}  \mathbf{a}_{\rm RIS}(\psi_{\rm R}^q, \omega^q) \mathbf{a}_{2}^{\rm H}(\psi_{\rm T}^q),
\end{equation}
where each $\beta_{q} $ is the complex-valued gain of the $q$-th UE$_2$-RIS channel component with $\psi_{\rm R}^q$ and $\omega^q$ denoting its azimuth AoA from the RIS and azimuth AoD at the UE$_2$, respectively, and $\psi_{\rm T}^q$ is the elevation AoA at the RIS. Finally, the $N_2\times1$ steering vector at the UE$_2$ is given similar to \eqref{eq:steering_UE1} as: 
\begin{align}
    \mathbf{a}_{2}(\psi) &\triangleq \boldsymbol{\beta}(\cos \psi, N_2).
\end{align}

\section{Channel Estimation at the Receiving RIS Side}\label{sec:Channel_Estimation}
As previously discussed, the outputs $\mathbf{y}_{\rm RF}(t)$ of the $N_{\rm RF}$ RX RF chains at the receiving RIS at each $t$-th training time slot, which are given by expression~\eqref{eq:received_RIS_RF_chains} for a given absorption phase configuration $\mathbf{U}$, feed the RIS baseband processor, and will be used for estimating the channel matrices $\mathbf{H}_1$ and $\mathbf{H}_2$, as follows. 


Recall that the $N_{\rm RIS} \times N_{\rm RF}$ matrix $\mathbf{U}$, including the elements $u_{i,n}$ $\forall$$i,n$ chosen from $\mathcal{W}_{\rm a}$ in~\eqref{eq: code}, represents one possible absorption phase profile for the overall semi-passive RIS (i.e., one per $N_{\rm E}$ group of metamaterials). We also introduce the matrix $\mathbf{W}\in \mathbb{C}^{N_{\rm RIS} \times N_{\rm RIS}}$ which is composed of $N_{\rm E} \left(=N_{\rm RIS}/N_{\rm RF}\right)$ distinct absorption profiles, chosen from the $|\mathcal{W}|^{N_{\rm E}}$ available ones per $N_{\rm E}$-element group, concatenated column-wise (i.e., horizontal concatenation of distinct $\mathbf{U}$ matrices in~\eqref{eq:Us}). Let also $\boldsymbol{\omega}(t) \in \{0,1\}^{N_{\rm RIS} \times 1}$ be an $N_{\rm RIS}$-element column-vector with $N_{\rm RIS}-N_{\rm RF}$ zeros and $N_{\rm RF}$ unity elements, where the positions of the latter elements are chosen at each $t$-th time interval according to the uniform distribution over the set $\{1,2,\ldots,N_{\rm RIS}\}$. Clearly, the operation $\boldsymbol{\omega}(t) \circ \left(\mathbf{W}^{\rm T} \mathbf{y}_{\rm RIS}(t)\right)$, with $\mathbf{y}_{\rm RIS}(t)=\mathbf{P}^{\rm H} \left(\mathbf{H}_1 \mathbf{s}_1(t) + \mathbf{H}_2 \mathbf{s}_2(t)\right)$ from \eqref{eq:received_RIS_RF_chains}, implies random selection of $N_{\rm RF}$ columns of $\mathbf{W}^{\rm T} \mathbf{y}_{\rm RIS}(t)$, i.e., the $N_{\rm RF}$ outputs of the RX RF chains using $N_{\rm RF}$ absorption phase profiles from the $N_{\rm E}$ available ones in $\mathbf{W}$. To this end, expression~\eqref{eq:received_RIS_RF_chains} can be rewritten to indicate random selection of absorption profiles for all $N_{\rm E}$-element groups in the following zero-padded form:
\begin{align}\label{eq:r_t}
     \mathbf{r}(t) \triangleq \boldsymbol{\omega}(t) \circ \left(\mathbf{W}^{\rm T} \mathbf{y}_{\rm RIS}(t)\right)+\mathbf{n}(t),
\end{align}
where for the $N_{\rm RF}$ non-zero elements of $\mathbf{r}(t)\in \mathbb{C}^{N_{\rm RIS} \times 1}$ and $\mathbf{n}(t)\in \mathbb{C}^{N_{\rm RIS} \times 1}$ holds that $[\mathbf{r}(t)]_i = [\mathbf{y}_{\rm RF}(t)]_i$ and $[\mathbf{n}(t)]_i = [\mathbf{n}_{\rm RF}(t)]_i$ $\forall$$i$, respectively.  

\subsection{Estimation Problem Formulation}
Using~\eqref{eq:r_t} for $T$ consecutive training time slots, the following expression for the $T$ received pilot signals at the outputs of the $N_{\rm RF}$ RX RF chains of the receiving RIS, which then become available at the RIS baseband processor, is deduced (note that, for each of the $T$ columns in the following matrix $\mathbf{R} \in \mathbb{C}^{N_{\rm RIS} \times T}$, only $N_{\rm RF}$ elements from the $N_{\rm RIS}$ ones have nonzero values):
\begin{equation}\label{eq:linear_system}
    \mathbf{R} \triangleq \mathbf{\Omega} \circ \left(\mathbf{W}^{\rm T} \mathbf{Y}_{\rm RIS}\right) + \mathbf{N},
\end{equation}
where $\mathbf{Y}_{\rm RIS}\triangleq[\mathbf{y}_{\rm RIS}(1)\,\mathbf{y}_{\rm RIS}(2)\,\cdots\,\mathbf{y}_{\rm RIS}(T)] \in \mathbb{C}^{N_{\rm RIS} \times T}$, $\mathbf{\Omega}\triangleq[\boldsymbol{\omega}(1)\,\boldsymbol{\omega}(2)\cdots\, \boldsymbol{\omega}(T)]$ is an $N_{\rm RIS} \times T$ matrix having $N_{\rm RF}T$ unity elements, and the complex-valued $N_{\rm RIS} \times T$ matrix $\mathbf{N}$ is defined as $\mathbf{N}\triangleq [\mathbf{n}(1)\,\mathbf{n}(2)\,\cdots\,\mathbf{n}(T)]$. Note that~\eqref{eq:linear_system} is a linear system of equations including the $N_\text{RIS} (N_1 + N_2)$ unknown elements of matrices $\mathbf{H}_1$ and $\mathbf{H}_2$ that need to be recovered via explicit channel estimation. To this end, conventional model-agnostic estimation approaches require the utilization of $T > N_\text{RIS} (N_1 + N_2)$ training time slots. However, exploiting the potential low rank and sparsity of the involved channel matrices, the training requirements may decrease significantly~\cite{MAD_TWC_2023,Vlachos2019WidebandSampling,Vlachos_SPL2018}. On this premise, we employ the following beamspace representation~\cite{6484896} for the unknown channel matrices $\mathbf{H}_1$ and $\mathbf{H}_2$: 
\begin{align}
\mathbf{H}_1 &= \mathbf{D}_{\rm RIS} \mathbf{Z}_1 \mathbf{\tilde{D}}_{1}^{\rm H}, \\
\mathbf{H}_2 &= \mathbf{D}_{\rm RIS} \mathbf{Z}_2 \mathbf{\tilde{D}}_{2}^{\rm H},
\end{align}
where $\mathbf{D}_{\rm RIS} \in \mathbb{C}^{N_{\rm RIS} \times N_{\rm RIS}}$, $\mathbf{\tilde{D}}_1 \in \mathbb{C}^{N_1 \times N_1}$, and $\mathbf{\tilde{D}}_2 \in \mathbb{C}^{N_2 \times N_2}$ represent Discrete Fourier Transform (DFT) matrices, while $\mathbf{Z}_1\in \mathbb{C}^{N_{\rm RIS} \times N_1}$ and $\mathbf{Z}_2\in \mathbb{C}^{N_{\rm RIS} \times N_2}$ indicate the beamspace representations of $\mathbf{H}_1$ and $\mathbf{H}_2$, respectively. We also define the extended channel matrix $\mathbf{\bar{H}}\triangleq[\mathbf{H}_1\,\mathbf{H}_2]\in \mathbb{C}^{N_{\rm RIS} \times N_{\rm UEs}}$, with $N_{\rm UEs} \triangleq N_1+N_2$, in the beamspace as follows:
\begin{equation}
\mathbf{\bar{H}} = \mathbf{D}_{\rm RIS}\mathbf{\bar{Z}} \mathbf{D}_{\text{UEs}}^{\text{H}},
\end{equation}
where $\mathbf{\bar{Z}} \triangleq \left[\mathbf{Z}_1\,\mathbf{Z}_2\right]$ is the concatenation of the beamspace matrices, and $\mathbf{D}_{\text{UEs}} \triangleq \text{diag}\left(\mathbf{\tilde{D}}_1, \mathbf{\tilde{D}}_2\right)$ indicating a block diagonal matrix. 

Putting all above together, the beamspace representation of the unknown extended channel matrix $\mathbf{\bar{H}}$ can be incorporated into the linear system of equations in~\eqref{eq:linear_system} to express the impinging pilot signals on the RIS unit elements as follows:
\begin{equation}\label{eq:input_output_1}
    \mathbf{Y}_{\rm RIS} = \mathbf{P}^{\rm H}\mathbf{D}_\text{RIS}\mathbf{\bar{Z}} \mathbf{D}_\text{UEs}^{\rm H} \mathbf{\bar{S}}, 
\end{equation}
where $\mathbf{\bar{S}} \triangleq \left[ \begin{array}{cc} \mathbf{S}_1^{\rm T} & \mathbf{S}_2^{\rm T} \end{array} \right]^{\rm T} \in \mathbb{C}^{N_{\rm UEs} \times T}$ with $\mathbf{S}_1\in \mathbb{C}^{N_1 \times T}$ and $\mathbf{S}_2\in \mathbb{C}^{N_2 \times T}$ denoting the pilot symbols transmitted from UE$_1$ and UE$_2$, respectively.

We now focus on exploiting the low rank and sparsity of $\mathbf{\bar{H}}$ for its estimation, and formulate the following problem for obtaining $\hat{\mathbf{H}}_1$ and $\hat{\mathbf{H}}_2$:
\begin{align}
\mathcal{OP}_2: \min_{\mathbf{\bar{H}}, \mathbf{\bar{Z}}} & \,\, \tau_Y \left\Vert \mathbf{\bar{H}} \right\Vert_* + \tau_Z \left\Vert \mathbf{\bar{Z}} \right\Vert_1 + \frac{1}{2} \left\Vert \mathbf{R} - \mathbf{\Omega} \circ \mathbf{W}^{\rm T} \mathbf{P}^{\rm H}\mathbf{\bar{H}} \mathbf{\bar{S}} \right\Vert_{\rm F}^2 \nonumber \\
\textrm{s.t.} &\,\,\mathbf{\bar{H}} = \mathbf{D}_\text{RIS} \mathbf{\bar{Z}} \mathbf{D}_\text{UEs}^{\rm H},\nonumber
\end{align}
where $\mathbf{\bar{H}}$'s nuclear norm in the objective function imposes its low rank property, whereas the $\ell_1$-norm of $\mathbf{\bar{Z}}$ enforces its sparse structure. In addition, the weighting factors $\tau_Y,\tau_Z>0$ depend, in general, on the number of the distinct $P$ and $Q$ MIMO channel propagation paths.  

\subsection{An ADMM-Based Channel Estimation Algorithm}
The $\mathcal{OP}_2$ can be efficiently solved via the ADMM algorithm. To this end, the Lagrangian function of this problem can be obtained as follows:
\begin{align}
\mathcal{L}_\rho(\mathbf{\bar{H}}, \mathbf{\bar{Z}}, \mathbf{\Gamma}) =& \,\,\tau_Y  \left\Vert \mathbf{\bar{H}} \right\Vert_* + \tau_Z \left\Vert \mathbf{\bar{Z}} \right\Vert_1 + \frac{1}{2} \left\Vert \mathbf{R} - \mathbf{\Omega} \circ \mathbf{W}^{\rm T} \mathbf{P}^{\rm H}\mathbf{D}_\text{RIS} \mathbf{\bar{Z}} \mathbf{D}_\text{UEs}^{\rm H} \mathbf{\bar{S}} \right\Vert_{\rm F}^2 \nonumber 
\\& + \text{tr}\left(\mathbf{\Gamma}^{\rm H} \left(\mathbf{\bar{H}} - \mathbf{D}_\text{RIS} \mathbf{\bar{Z}} \mathbf{D}_\text{UEs}^{\rm H}\right) \right) + \text{tr}\left(\left(\mathbf{\bar{H}} - \mathbf{D}_\text{RIS} \mathbf{\bar{Z}} \mathbf{D}_\text{UEs}^{\rm H}\right)^{\rm H} \mathbf{\Gamma} \right) \nonumber \\&+ \frac{\rho}{2} \left\Vert \mathbf{\bar{H}} - \mathbf{D}_\text{RIS} \mathbf{\bar{Z}} \mathbf{D}_\text{UEs}^{\rm H} \right\Vert_{\rm F}^2,
\label{eq:Lagrangian_1}
\end{align}
where $\mathbf{\Gamma} \in \mathbb{C}^{N_{\rm RIS} \times N_{\rm UEs}}$ is the dual variable for both UE$_1$ and UE$_2$, and $\rho\in(0,1)$ denotes the ADMM's stepsize. In particular, following the ADMM methodology, at each $\ell$-th algorithmic iteration ($\ell=1,2, \ldots$), the following separate sub-problems need to be solved:
\begin{align}
    \mathbf{\bar{H}}^{(\ell+1)} &= \arg \min_{\mathbf{\bar{H}}} \mathcal{L}_\rho\left(\mathbf{\bar{H}}, \mathbf{\bar{Z}}^{(\ell)}, \mathbf{\Gamma}^{(\ell)}\right),  \label{eq:admm_problem_step_H} \\
     \mathbf{\bar{Z}}^{(\ell+1)} &= \arg\min_{\mathbf{\bar{Z}}} \mathcal{L}_\rho\left(\mathbf{\bar{H}}^{(\ell+1)}, \mathbf{\bar{Z}}, \mathbf{\Gamma}^{(\ell)}\right),  \label{eq:admm_problem_step_Z} \\
    \mathbf{\Gamma}^{(\ell+1)} &= \mathbf{\Gamma}^{(\ell)} + \rho \left(\mathbf{\bar{H}}^{(\ell+1)} - \mathbf{D}_\text{RIS}\mathbf{\bar{Z}}^{(\ell+1)} \mathbf{D}_\text{UEs}^{\rm H}\right), \label{eq:admm_problem_dual_Gamma}
\end{align}
where $\mathbf{\bar{Z}}^{(1)} = \mathbf{0}_{N_{\rm RIS} \times N_{\rm UEs}}$ and $\mathbf{\Gamma}^{(1)} = \mathbf{0}_{N_{\rm RIS} \times N_{\rm UEs}}$.

\subsubsection{Minimization over $\mathbf{\bar{H}}$} Eliminating the terms of the Langrangian function that do not depend on $\mathbf{\bar{H}}$, results in the following formulation:
\begin{align}
    \min_{\mathbf{\bar{H}}}\,\, & \tau_Y \left\Vert 
    \mathbf{\bar{H}} \right\Vert_* +\text{tr}\left(\left(\mathbf{\Gamma}^{(\ell)}\right)^{\rm H} \left(\mathbf{\bar{H}} - \mathbf{D}_\text{RIS} \mathbf{\bar{Z}}^{(\ell)} \mathbf{D}_\text{UEs}^{\rm H}\right) \right) \nonumber 
    \\ & +\text{tr}\left(\left(\mathbf{\bar{H}} - \mathbf{D}_\text{RIS} \mathbf{\bar{Z}}^{(\ell)} \mathbf{D}_\text{UEs}^{\rm H}\right)^{\rm H} \mathbf{\Gamma}^{(\ell)}\right)+
    \frac{\rho}{2} \left\Vert \mathbf{\bar{H}} - \mathbf{D}_\text{RIS} \mathbf{\bar{Z}}^{(\ell)} \mathbf{D}_\text{UEs}^{\rm H} \right\Vert_{\rm F}^2.
\end{align}
Let us introduce the weight parameters $\tau_{Y}$ and $\tau_{Z}$  that depend on the number of propagation paths \cite{Vlachos2019WidebandSampling}. Then, completing the square term by adding the term $\frac{2 }{\rho} \Vert \mathbf{\Gamma}^{(\ell)} \Vert^2_F$, we have the equivalent problem:
\begin{equation}\label{eq:opt_H_k}
    \min_{\mathbf{\bar{H}}}\,\, \frac{\tau_Y}{\rho} \Vert 
    \mathbf{\bar{H}} \Vert_* + \frac{\rho}{2} \left\Vert \mathbf{\bar{H}} - \left(\mathbf{D}_\text{RIS} \mathbf{\bar{Z}}^{(\ell)} \mathbf{D}_\text{UEs}^{\rm H} - \frac{2}{\rho}\mathbf{\Gamma}^{(\ell)}\right) \right\Vert_{\rm F}^2.
\end{equation}
\\
The optimization \eqref{eq:opt_H_k} can be solved efficiently via the SVT algorithm \cite{cai2010} parameterized by $\rho$, i.e., for each $\ell$-th algorithmic iteration:
\begin{equation}
    \mathbf{\bar{H}}^{(\ell+1)} = \text{SVT}_{\tau_Y/\rho}\left( \mathbf{D}_\text{RIS} \mathbf{\bar{Z}}^{(\ell)} \mathbf{D}_\text{UEs}^{\rm H} - \frac{2}{\rho}\mathbf{\Gamma}^{(\ell)} \right).
\end{equation}

\subsubsection{Minimization over $\mathbf{\bar{Z}}$} 
In a similar manner, keeping only the terms related with the $\mathbf{\bar{Z}}$ variable in \eqref{eq:Lagrangian_1}, yields the following formulation:
\begin{align}
\min_{\mathbf{\bar{Z}}}\,\, &\tau_Z \left\Vert \mathbf{\bar{Z}} \right\Vert_1 
+ \frac{1}{2} \left\Vert \mathbf{R} - \mathbf{\Omega} \circ \mathbf{W}^{\rm T} \mathbf{P}^{\rm H}\mathbf{D}_\text{RIS} \mathbf{\bar{Z}} \mathbf{D}_\text{UEs}^{\rm H} \mathbf{\bar{S}} \right\Vert_{\rm F}^2 \nonumber \\
&+ \text{tr}\left(\left(\mathbf{\Gamma}^{(\ell)}\right)^{\rm H} \left(\mathbf{\bar{H}}^{(\ell+1)} - \mathbf{D}_\text{RIS} \mathbf{\bar{Z}} \mathbf{D}_\text{UEs}^{\rm H}\right) \right) \nonumber \\ &
+ \text{tr}\left(\left(\mathbf{\bar{H}}^{(\ell+1)} - \mathbf{D}_\text{RIS} \mathbf{\bar{Z}} \mathbf{D}_\text{UEs}^{\rm H}\right)^{\rm H} \left(\mathbf{\Gamma}^{(\ell)}\right) \right) \nonumber \\ &+\frac{\rho}{2} \left\Vert \mathbf{\bar{H}}^{(\ell+1)} - \mathbf{D}_\text{RIS} \mathbf{\bar{Z}} \mathbf{D}_\text{UEs}^{\rm H} \right\Vert_{\rm F}^2. 
\end{align}
As previously, completing the square term by adding the term $\frac{2}{\rho} \Vert \mathbf{\Gamma}^{(\ell)} \Vert_{\rm F}^2$, results in the reformulation:
\begin{align}
 \min_{\mathbf{\bar{Z}}}\,\, &\tau_Z \left\Vert \mathbf{\bar{Z}} \right\Vert_1 
+ \frac{1}{2} \left\Vert \mathbf{R} - \mathbf{\Omega} \circ \mathbf{W}^{\rm T} \mathbf{P}^{\rm H} \mathbf{D}_\text{RIS} \mathbf{\bar{Z}} \mathbf{D}_\text{UEs}^{\rm H} \mathbf{\bar{S}} \right\Vert_{\rm F}^2 \nonumber \\
&+ \frac{\rho}{2} \left\Vert \mathbf{\bar{H}}^{(\ell+1)} - \frac{2}{\rho} \mathbf{\Gamma}^{(\ell)} -  \mathbf{D}_\text{RIS} \mathbf{\bar{Z}} \mathbf{D}_\text{UEs}^{\rm H}\right\Vert_{\rm F}^2. \label{eq:min_Z_2}
\end{align}
To proceed, let us express the problem in its vectorized form. In particular, the second term in \eqref{eq:min_Z_2} can be written as follows:
\begin{align}
&\left\Vert \mathbf{R} - \mathbf{\Omega} \circ \mathbf{W}^{\rm T} \mathbf{P}^{\rm H}\mathbf{D}_\text{RIS} \mathbf{\bar{Z}} \mathbf{D}_\text{UEs}^{\rm H} \mathbf{\bar{S}} \right\Vert_{\rm F}^2 \nonumber \\ 
&= \left\Vert \text{vec}(\mathbf{R}) - \text{vec}(\mathbf{\Omega} \circ \mathbf{W}^{\rm T} \mathbf{P}^{\rm H} \mathbf{D}_\text{RIS} \mathbf{\bar{Z}} \mathbf{D}_\text{UEs}^{\rm H} \mathbf{\bar{S}}) \right\Vert_2^2 \nonumber \\ 
&= \left\Vert \text{vec}(\mathbf{R}) - \text{diag}(\text{vec}(\mathbf{\Omega})) \text{vec}(\mathbf{W}^{\rm T} \mathbf{P}^{\rm H} \mathbf{D}_\text{RIS} \mathbf{\bar{Z}} \mathbf{D}_\text{UEs}^{\rm H} \mathbf{\bar{S}}) \right\Vert_2^2 \nonumber \\ 
&= \left\Vert \underbrace{\text{vec}(\mathbf{R})}_{\triangleq\boldsymbol{\xi}_1} - \underbrace{\text{diag}(\text{vec}(\mathbf{\Omega})) \left( \left(\mathbf{D}_\text{UEs}^{\rm H} \mathbf{\bar{S}}\right)^{\rm T} \otimes \mathbf{W}^{\rm T}\mathbf{P}^{\rm H} \mathbf{D}_\text{RIS} \right)}_{\triangleq\mathbf{\Phi}_1} \underbrace{\text{vec}(\mathbf{\bar{Z}})}_{\triangleq\mathbf{\bar{z}}} \right\Vert_2^2, 
\end{align}
and the third term as:
\begin{align}
& \left\Vert \mathbf{D}_\text{RIS} \mathbf{\bar{Z}} \mathbf{D}_\text{UEs}^{\rm H} - \left(\mathbf{\bar{H}}^{(\ell+1)} + \frac{2}{\rho} \mathbf{\Gamma}^{(\ell)}\right)\right\Vert_{\rm F}^2 \nonumber \\
&= \left\Vert \text{vec}\left(\mathbf{D}_\text{RIS} \mathbf{\bar{Z}} \mathbf{D}_\text{UEs}^{\rm H}\right) - \underbrace{\text{vec}\left(\mathbf{\bar{H}}^{(\ell+1)} + \frac{2}{\rho} \mathbf{\Gamma}^{(\ell)}\right)}_{\triangleq\boldsymbol{\xi}_2}\right\Vert_2^2 = \left\Vert \underbrace{\left(\mathbf{D}_\text{UEs}^* \otimes \mathbf{D}_\text{RIS}\right)}_{\triangleq\mathbf{\Phi}_2} \mathbf{\bar{z}}- \boldsymbol{\xi}_2 \right\Vert_2^2.
\end{align}
Then, problem \eqref{eq:min_Z_2} becomes:
\begin{equation}
\min_{\mathbf{\bar{z}}}\,\, \tau_Z \left\Vert \mathbf{\bar{z}} \right\Vert_1 
+ \frac{1}{2} \left\Vert \boldsymbol{\xi}_1 - \mathbf{\Phi}_1\mathbf{\bar{z}} \right\Vert_2^2 + \frac{\rho}{2}\left\Vert \mathbf{\Phi}_2 \mathbf{\bar{z}} - \boldsymbol{\xi}_2 \right\Vert_2^2, \label{LassoNew} 
\end{equation}
or equivalently, by setting $\boldsymbol{\bar{\xi}} \triangleq \left[ \begin{array}{cc} \boldsymbol{\xi}_1^{\rm T} & \sqrt{\rho}\boldsymbol{\xi}_2^{\rm T} \end{array} \right]^{\rm T}$ and $\mathbf{\bar{\Phi}} \triangleq \left[ \begin{array}{cc} \mathbf{\Phi}_1^{\rm T} & \sqrt{\rho}\mathbf{\Phi}_2^{\rm T} \end{array} \right]^{\rm T}$, as it can be written as follows:
\begin{equation}
\min_{\mathbf{\bar{z}}}\,\, \tau_Z \left\Vert \mathbf{\bar{z}} \right\Vert_1 
+ \frac{1}{2} \left\Vert \boldsymbol{\bar{\xi}} - \mathbf{\bar{\Phi}}\mathbf{\bar{z}} \right\Vert_2^2 , \label{lasso}
\end{equation}
which expresses the well-known LASSO problem \cite{Tibshirani1996}.
This problem can be solved via the soft-thresholding operator $\mathcal{S}_{\tau_Z}$, where the solution is given by:
\begin{equation}
\mathbf{\bar{z}}^{(\ell+1)} = \mathcal{S}_{\tau_Z}\left(\mathbf{\bar{\Phi}}^\dagger \boldsymbol{\bar{\xi}}\right),
\end{equation}
 where $\mathbf{\bar{\Phi}}^\dagger$ represents the Moore–Penrose pseudoinverse of $\mathbf{\bar{\Phi}}$ and the operator $\mathcal{S}_{\tau_Z}(\cdot)$ is defined as follows:
\begin{align}
    \mathcal{S}_{\tau_Z}(\cdot) \triangleq &\text{sgn}\left\{\texttt{Re}(\cdot)\right\} \circ \max \left\{\vert \mathtt{Re}(\cdot) \vert - \tau_Z, 0\right\}
    \nonumber \\
    &+ \text{sgn}\left\{\texttt{Im}(\cdot)\right\} \circ \max \left\{\vert \mathtt{Im}(\cdot) \vert - \tau_Z, 0\right\}.
\end{align}

The overall approach for solving $\mathcal{OP}_2$, i.e., the proposed estimation of the channel matrices $\mathbf{H}_1$ and $\mathbf{H}_2$ per channel coherence time, is summarized in Algorithm~\ref{algorithm:proposed_algorithm_1}. Step $2$ focuses on the minimization of $\mathcal{L}_\rho(\cdot)$ in~\eqref{eq:Lagrangian_1} over $\mathbf{\bar{H}}$, whereas Step $3$ performs the minimization of the same function with respect to $\mathbf{\bar{Z}}$. Finally, in Step $4$, the dual variables for both UE$_1$ and UE$_2$ included in matrix $\mathbf{\Gamma}$ are updated.

\begin{algorithm}[!t]
	\caption{Proposed Estimation for $\mathbf{\bar{H}}\triangleq[\mathbf{H}_1\,\mathbf{H}_2]$}
	\begin{algorithmic}[1]
		\REQUIRE     $\mathbf{R}$, $\mathbf{\bar{\Phi}}^{\dagger}$, $\mathbf{\Gamma}^{(1)} = \mathbf{0}$, $\mathbf{\bar{Z}}^{(1)} = \mathbf{0}$,  $\tau_Z$, and $I_{\rm max}$.
		\ENSURE $\mathbf{\bar{H}}^{(I_{\rm max}+1)}$
		\FOR {$\ell=1,2,\ldots,I_{\rm max}$}     

        \STATE Set $\mathbf{\bar{H}}^{(\ell+1)} = \text{SVT}_\rho\left(\left(\mathbf{D}_\text{RIS} \mathbf{\bar{Z}}^{(\ell)}\right) \mathbf{D}_\text{UEs}^{\rm H} - \frac{4}{\rho}\mathbf{\Gamma}^{(\ell)} \right)$. \\
        \STATE Set $\mathbf{w} = \text{vec}\left(\mathbf{\bar{\Phi}}^\dagger \, \text{vec}\left( \mathbf{R} - \left(\mathbf{\bar{H}}^{(\ell+1)} + \frac{4}{\rho} \mathbf{\Gamma}\right)\right)\right)$. \\
        \STATE Compute:\\ $\mathbf{\bar{z}} = \text{sgn}\{\texttt{Re}(\mathbf{w})\} \circ  \max \{\vert \mathtt{Re}(\mathbf{w}) \vert - \tau_Z, 0\} + \text{sgn}\{\texttt{Im}(\mathbf{w})\} \circ \max \{\vert \mathtt{Im}(\mathbf{w}) \vert - \tau_Z, 0\}$ \\ and reconstruct the matrix $\mathbf{\bar{Z}}^{(\ell+1)} = \text{unvec}(\mathbf{\bar{z}})$. \\
        
        \STATE Set $\mathbf{\Gamma}^{(\ell+1)} = \mathbf{\Gamma}^{(\ell)} + \rho \left(\mathbf{\bar{H}}^{(\ell+1)} - \mathbf{D}_\text{RIS}\mathbf{\bar{Z}}^{(\ell+1)} \mathbf{D}_\text{UEs}^{\rm H}\right)$.
        \ENDFOR
	\end{algorithmic}
	\label{algorithm:proposed_algorithm_1}
\end{algorithm}

\subsubsection{Computational Complexity}
In this subsection, the computational complexity of Algorithm~\ref{algorithm:proposed_algorithm_1} is analyzed and a technique to enhance its implementation efficiency is presented.

The pseudoinverse $\mathbf{\bar{\Phi}}^\dagger$ is one of the inputs of Algorithm~\ref{algorithm:proposed_algorithm_1} requiring complexity of the order of $\mathcal{O}((N_{\rm RIS}^2 N_{\rm UEs} (T+N_{\rm UEs}))^3)$. It will be next shown that this complexity can be significantly reduced by exploiting the special structure of matrix $\mathbf{\bar{\Phi}}$. The algorithm operates iteratively executing $I_{\rm max}$ iterations, hence, its complexity is directly related to this iteration count. The complexity of each individual iteration is analyzed as follows:
\\
\begin{itemize}
    \item The SVT algorithm included in Step $2$ requires $\mathcal{O}\left(N_{\rm UEs} N_{\rm RIS}^2\right)$ operations; note that $N_{\rm RIS}$ is much larger than the number of $N_{\rm UEs}$. Essentially, its complexity is linked with the complexity of the Singular Value Decomposition (SVD). Although standard SVD algorithms can be computationally expensive, several efficient approaches have been developed to improve performance, particularly for large-scale problems. Those include randomized SVD, Lanczos-based methods, and iterative techniques (e.g., power iteration), reaching a complexity of the order of $\mathcal{O}\left(N_{\rm RIS} N_{\rm UEs} \min\left\{N_{\rm RIS}, N_{\rm UEs}\right\}\right)$.
    \item Step $3$ requires the calculation of matrix $\mathbf{w}$, with complexity related to the matrix-vector multiplication $\mathbf{\bar{\Phi}}^\dagger\text{vec}\left( \mathbf{R} \!-\! \left(\mathbf{\bar{H}}^{(\ell+1)} \!+\! \frac{4}{\rho} \mathbf{\Gamma}\right)\right)$, resulting in $\mathcal{O}\left(\left(T+N_{\rm UEs}\right) N_{\rm RIS} N_{\rm UEs} \right)$ operations.
    \item Step 4 computes the vector $\mathbf{\bar{z}}$ with $\mathcal{O}\left(N_{\rm RIS} (T+N_{\rm UEs})\right)$ complexity needed for the $\max\{\cdot\}$ operator applied on vector $\mathbf{w}$.
    \item Finally, Step 5 requires complexity of the order of $\mathcal{O}\left(N_{\rm RIS}^2 N_{\rm UEs} \right) + \mathcal{O}\left(N_{\rm UEs}^2 N_{\rm RIS} \right)$ for the matrix-matrix multiplications $\mathbf{D}_\text{RIS}\mathbf{\bar{Z}}^{(\ell+1)} \mathbf{D}_\text{UEs}^{\rm H}$.
\end{itemize}

The matrix $\mathbf{\bar{\Phi}}^\dagger$ in Algorithm~\ref{algorithm:proposed_algorithm_1} is essential for solving the LASSO problem in~\eqref{LassoNew}. Since it can be precomputed before the iterations and remains unchanged throughout, we outline a more efficient approach that leverages its unique structure. Recall that $N_{\rm RIS} (T + N_{\rm UEs}) \times N_{\rm RIS} N_{\rm UEs}$ matrix $\mathbf{\bar{\Phi}}$ is defined as $\mathbf{\bar{\Phi}} \triangleq \left[ \begin{array}{cc} \mathbf{\Phi}_1^{\rm T} & \sqrt{\rho}\mathbf{\Phi}_2^{\rm T} \end{array} \right]^{\rm T}$, where:
\begin{eqnarray*}
\mathbf{\Phi}_{1} &= & \text{diag}\left(\text{vec}\left(\mathbf{\Omega} + \frac{\rho}{2} \mathbf{I}\right)\right) \left(\underbrace{\mathbf{\bar{S}}^{\rm T} \mathbf{D}_{\rm UEs}^\text{*} \otimes \mathbf{W}^{\rm H} \mathbf{D}_{\rm RIS}}_{\triangleq\mathbf{\Phi}'_{1}}\right), \label{eq:pseudoinv1} \\
\mathbf{\Phi}_{2} &= &  \mathbf{D}_{\rm UEs}^\text{*} \otimes  \mathbf{D}_{\rm RIS}  \label{eq:pseudoinv2},
\end{eqnarray*}
It can be easily concluded that $\mathbf{\bar{\Phi}}$ has a block diagonal structure, hence, its pseudoinverse is calculated as follows~\cite{Baksalary2007}:
\begin{equation}
\mathbf{\bar{\Phi}}^{\dagger}=
\begin{bmatrix}
\mathbf{\Phi}_{1}   \\
\sqrt{\rho}\mathbf{\Phi}_{2}   
\end{bmatrix}^{\dagger}=\left[ \left(  \mathbf{\Phi}_{1}\mathbf{P}_{\mathbf{\Phi}_{2}}^{\perp}\right)^{\dagger}  \left(  \sqrt{\rho}\mathbf{\Phi}_{2}\mathbf{P}_{\mathbf{\Phi}_{1}}^{\perp}\right)^{\dagger} \right].\label{eq:BLpseudoinv1}
\end{equation}
For general matrices $\mathbf{A}$ and $\mathbf{B}$, it holds that $\left(\mathbf{A}\mathbf{B}  \right)^{\dagger}=\mathbf{B}  ^{\dagger}\mathbf{A}^{\dagger}$ and matrix $\mathbf{P}_{\mathbf{A}}^{\perp}\triangleq \mathbf{I}-\mathbf{A}^{\dagger}\mathbf{A}$ represents the orthogonal projection into the kernel of $\mathbf{A}$~\cite{Baksalary2007}. To this end, \eqref{eq:BLpseudoinv1} can be re-written as follows:
\begin{equation}
\mathbf{\bar{\Phi}}^{\dagger}=\left[ \left(  \mathbf{I}-\mathbf{\Phi}_{2}^{\dagger}\mathbf{\Phi}_{2}\right)^{\dagger}\mathbf{\Phi}_{1}^{\dagger}\,\,\,   \frac{1}{\sqrt{\rho}}\left(  \mathbf{I}-\mathbf{\Phi}_{1}^{\dagger}\mathbf{\Phi}_{1}\right)^{\dagger}\mathbf{\Phi}_{2}^{\dagger} \right] \label{BLpseudoinv2}.
\end{equation}
To compute the pseudoinverse of $\mathbf{\Phi}_{1}^{'}$, we utilize the Kronecker product property, which holds for the Moore–Penrose pseudoinverse \cite{Langville2004}. Specifically: $$(\mathbf{\Phi}_{1}^{'})^\dagger = \left( \mathbf{\bar{S}}^{\rm T} \mathbf{D}_{\rm UEs}^\text{*} \right)^\dagger \otimes \left( \mathbf{W}^{\rm H} \mathbf{D}_{\rm RIS} \right)^\dagger =  \left(\mathbf{D}_{\rm UEs}^\text{*}\right)^\dagger \left(\mathbf{\bar{S}}^{\text{T}}\right)^\dagger \otimes \mathbf{D}_{\rm RIS} ^\dagger (\mathbf{W}^{\rm H})^\dagger.$$ 
In addition since the DFT matrices $\mathbf{D}_{\rm RIS}$ and $\mathbf{D}_{\rm UEs}^{\text{*}}$ are orthonormal, the following is deduced~\cite{Greville1966}: 
\begin{equation}
\left(\mathbf{\Phi}_{1}^{'}\right)^\dagger = \mathbf{D}_{\rm UEs}^{\rm T} (\mathbf{\bar{S}}^{\text{T}})^\dagger \otimes \mathbf{D}_{\rm RIS}^{\rm H}(\mathbf{W}^{\rm H})^\dagger. \label{eq:eff_lasso}
\end{equation}
The cost for the inversion of matrices $\mathbf{\bar{S}}$ and $\mathbf{W}$ is $\mathcal{O}(N_{\rm UEs}^2 T)$ and $\mathcal{O}(N_{\rm RIS}^3)$, respectively. Then, the cost for the Kronecker product is $\mathcal{O}(N_{\rm RIS}^2 N_{\rm UEs} T)$. 

Finally, the calculation of $\mathbf{\Phi}_{2}^{\dagger}$ in~\eqref{BLpseudoinv2} can be done efficiently. By using the property of the pseudoinverse of the Kronecker product~\cite{Langville2004} and the orthonormality of the DFT matrices, $\mathbf{\Phi}_{2}^{\dagger}$ is directly computed as follows:
\begin{equation}
    \left(\mathbf{\Phi}_{2}^{'}\right)^\dagger = \mathbf{D}_{\rm UEs}^{\rm T} \otimes \mathbf{D}_{\rm RIS}^{\rm H}
    \label{eq:eff_lasso2},
\end{equation}
which requires a computational complexity of $\mathcal{O}\left(N_{\rm RIS}^2 N_{\rm UEs}^2\right)$.

\section{Performance Evaluation Results and Discussion}\label{sec:Results}
This section presents and discusses performance evaluation results for both the proposed ADMM-based channel estimation approach detailed in Section~\ref{sec:Channel_Estimation} as well as the reflection phase configuration scheme presented in Section~\ref{sec:RIS_reflection_optimization}. Recall that these operations are both handled by the proposed semi-passive RIS hardware architecture and protocol in Section~\ref{sec:Model_Mode_or_Operation}, enabling signal reception and processing at the RIS side via the incorporation of a small number $N_{\rm RF}$ of RX RF chains and a baseband processing unit. 

Focusing first on the performance investigation of the proposed channel estimation approach, we have considered the typical Normalized Mean Squared Error (NMSE) metric, which is defined as follows:
\begin{equation}
    \text{NMSE} \triangleq \frac{1}{R}\sum_{r=1}^R \sum_{k=1}^2 \frac{\left\Vert \mathbf{H}_k - \mathbf{\hat{H}}_k \right\Vert_{\rm F}}{\left\Vert \mathbf{H}_k \right\Vert_{\rm F}},
\end{equation}
where $\mathbf{\hat{H}}_1$ and $\mathbf{\hat{H}}_2$ represent respectively the estimations for the RIS-UE$_1$ channel matrix $\mathbf{H}_1$ and the RIS-UE$_2$ channel matrix $\mathbf{H}_2$, while $R$ denotes the total number of Monte Carlo realizations used ($R\in[200,500]$ was used throughout all performance results). Apart from the presented Partially-Connected (PC) reception architecture for the semi-passive RIS, we have also considered, for comparison purposes, the more complex Fully-Connected (FC) reception architecture used in~\cite{AV_ICASSP_2020}, according to which all unit elements are connected to all RX RF chains~\cite{FD_MIMO_ISAC_Thz}.

\subsection{Convergence of ADMM-Based Channel Estimation}
\begin{figure}[!t]
    \centering
    \includegraphics[scale=0.65]{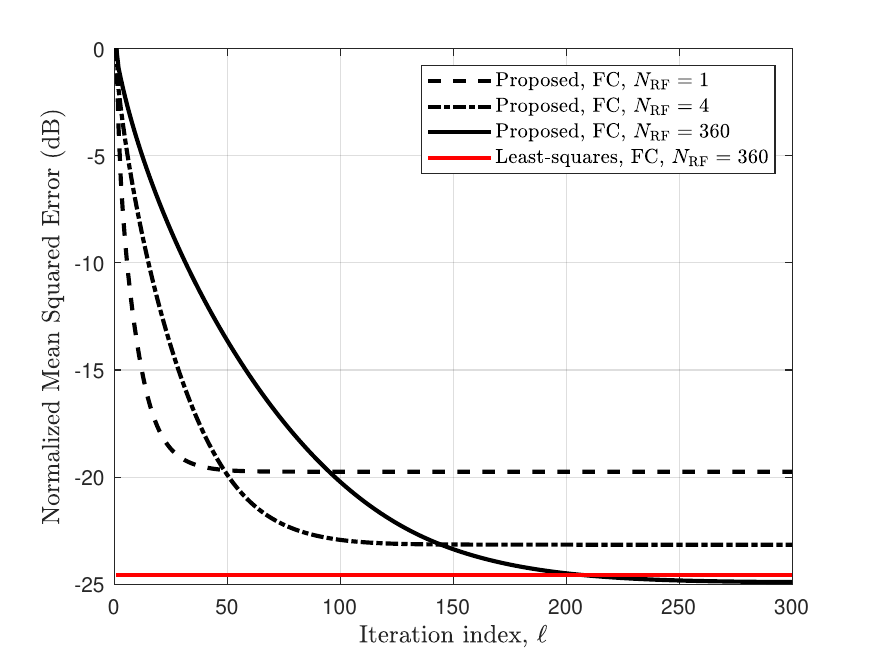}
    \caption{Convergence rate of Algorithm~\ref{algorithm:proposed_algorithm_1} for the ${\rm SNR}$ value $30$ dB, considering the parameters' setting: $N_{\rm RIS}=360$, $N_1=N_2=3$, $P=Q=1$, and $T=3240$. The number of the absorption phase levels per RIS unit element was set to $\vert \mathcal{W}_{\rm a} \vert = 9$.}
    \label{fig:convergence_30dB}
\end{figure}
\begin{figure}[!t]
    \centering
    \includegraphics[scale=0.65]{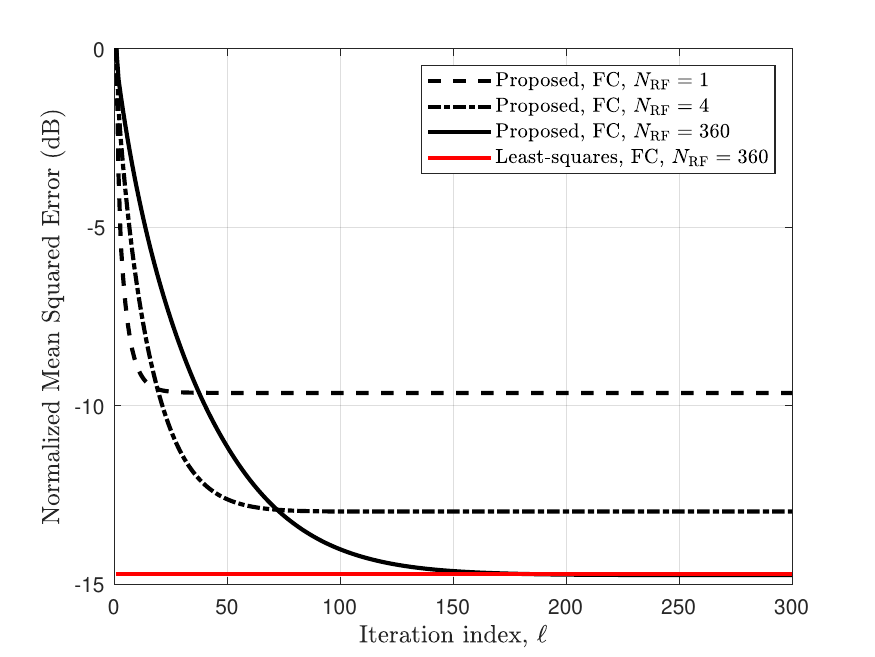}
    \caption{Same as in Fig.~\ref{fig:convergence_30dB}, but for the ${\rm SNR}$ value $10$ dB.}
    \label{fig:convergence_10dB}
\end{figure}
Figures~\ref{fig:convergence_30dB} and~\ref{fig:convergence_10dB} illustrate the convergence behavior of the proposed ADMM-based channel estimation Algorithm~\ref{algorithm:proposed_algorithm_1} for the ${\rm SNR}$ values $30$~dB and $10$~dB, respectively, considering a receiving RIS with $N_{\rm RIS}=360$ unit elements and three different versions of the FC reception architecture used in~\cite{AV_ICASSP_2020}, namely, with $N_{\rm RF}=\{1,4,360\}$ RX RF chains. Note that the $N_{\rm RF}=360$ case indicates the fully digital combining case. In fact, for this extreme case in terms of RIS hardware complexity, we have also simulated the least-squares channel estimation method for comparison purposes. In both figures, the number of the absorption phase levels per RIS unit element was set to $\vert \mathcal{W}_{\rm a} \vert = 9$, the numbers of antenna elements at UE$_1$ and UE$_2$ were $N_1=N_2=3$, the number of channel propagation paths for both links were set as $P=Q=1$, and $T=3240$ pilot symbols were used.

As observed from both Figs.~\ref{fig:convergence_30dB} and~\ref{fig:convergence_10dB}, the proposed iterative channel estimation approach for the fully digital combining case (i.e., for FC with $N_{\rm RF}=360$) achieves the accuracy of the least-squares estimator for the same combining architecture in a relatively small number of algorithmic iterations for both investigated ${\rm SNR}$ values, thus, validating our estimation framework. It can be also seen that, for FC with $N_{\rm RF}\ll360$, the estimation performance drops; this degradation becomes worse for lower ${\rm SNR}$ values and smaller number $N_{\rm RF}$ of RX RF chains. It is finally evident from both figures that the convergence of the proposed Algorithm~\ref{algorithm:proposed_algorithm_1} becomes faster for smaller $N_{\rm RF}$ values, though, to a lower NMSE point. 

\subsection{Channel Estimation Results}
\begin{figure}[!t]
    \centering
    \includegraphics[scale=0.65]{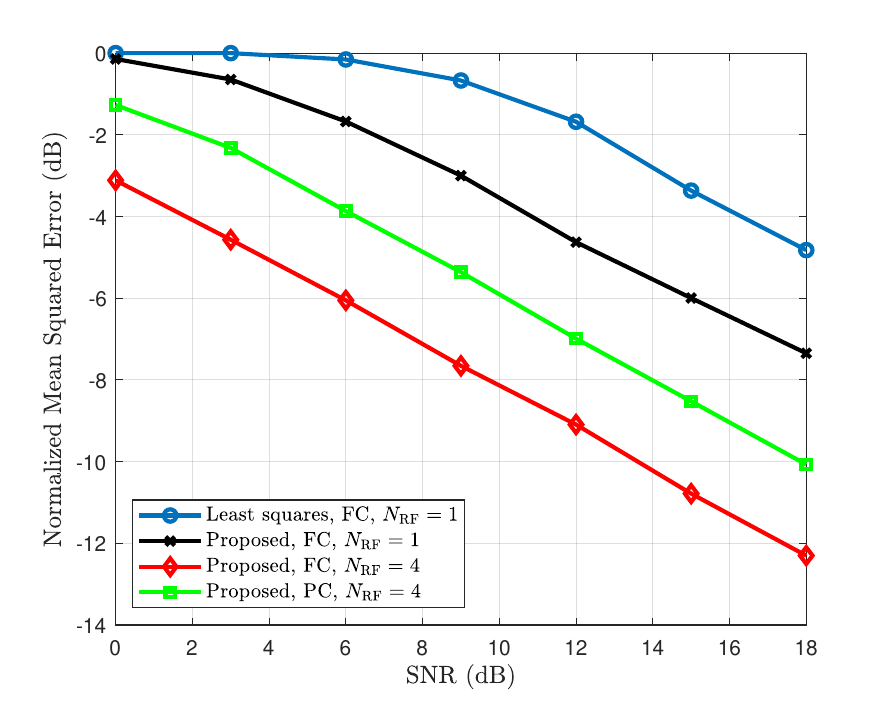}
    \caption{NMSE performance versus the ${\rm SNR}$ for the parameters' setting: $N_{\rm RIS}=256$, $N_1=N_2=3$, $P=Q=3$, and $T=2000$. Receiving RISs with $N_{\rm RF}=\{1,4\}$ RX RF chains have been considered, and the number of the absorption phase levels of each of their unit elements was set to $\vert \mathcal{W}_{\rm a} \vert = 5$.}
    \label{fig:mseVsnr_4RFchains}
\end{figure}
\begin{figure}[!t]
    \centering
    \includegraphics[scale=0.65]{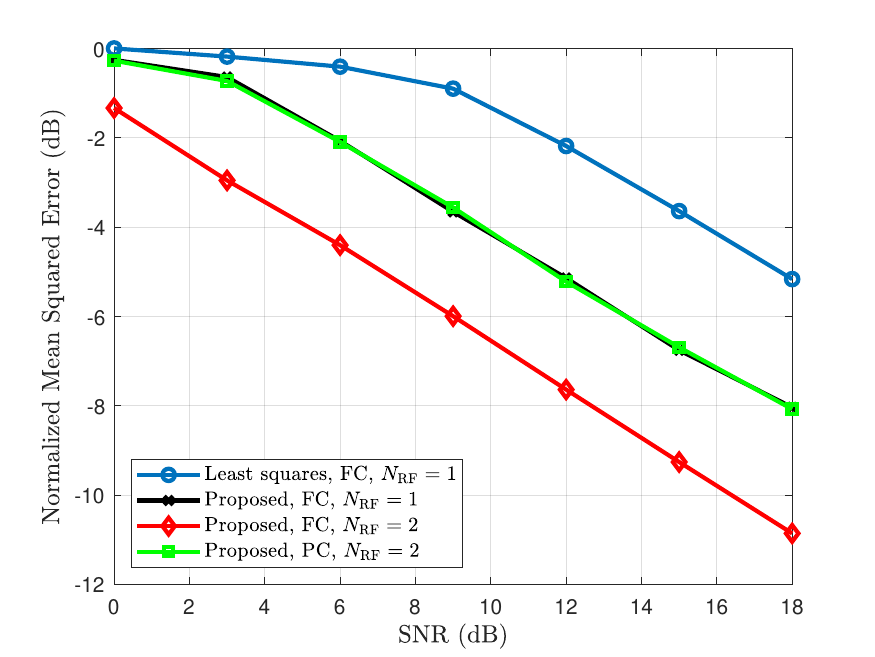}
    \caption{Same as in Fig.~\ref{fig:mseVsnr_4RFchains}, considering receiving RISs with $N_{\rm RF}=\{1,2\}$ RX RF chains.}
    \label{fig:mseVsnr_2RFchains}
\end{figure}

The NMSE performance of the proposed ADMM-based channel estimation Algorithm~\ref{algorithm:proposed_algorithm_1} as a function of the ${\rm SNR}$, considering a receiving RIS with $N_{\rm RIS}=256$ unit elements and different versions of the FC and PC reception architectures with $N_{\rm RF}=\{1,2,4\}$, is demonstrated in Figs.~\ref{fig:mseVsnr_4RFchains} and~\ref{fig:mseVsnr_2RFchains}. In both figures, the number of the absorption phase levels per RIS unit element was set to $\vert \mathcal{W}_{\rm a} \vert = 5$, the numbers of antenna elements at UE$_1$ and UE$_2$ were $N_1=N_2=3$, the number of channel propagation paths for both links were set as $P=Q=3$, and $T=2000$ pilot symbols were used. Least-squares channel estimation with a single-RF FC reception architecture has been also evaluated for comparison purposes.  

It can be observed from Fig.~\ref{fig:mseVsnr_4RFchains} that the least-squares method exhibits the poorest results, indicating that a solitary RF chain fails to provide sufficient information for effective convergence, even when trained with $T=2000$ pilot symbols. Conversely, it is shown in the figure that the NMSE of the proposed technique consistently decreases sufficiently as ${\rm SNR}$ increases, demonstrating improved accuracy across all investigated reception architecture settings. Clearly, the FC case with $N_{\rm RF}=4$ provides the best estimation performance over all ${\rm SNR}$ values. Then, the PC case with $N_{\rm RF}=4$ follows, strikes a balance between complexity and efficiency. Recall that the observed reduction in NMSE is attributable to this scheme's streamlined architecture, wherein each RX RF chain processes data from only a specific segment of the overall surface, optimizing performance while maintaining design simplicity. As shown, the worst NMSE channel estimation performance results from the application of the proposed Algorithm~\ref{algorithm:proposed_algorithm_1} to the FC case with $N_{\rm RF}=1$. Finally, Fig.~\ref{fig:mseVsnr_2RFchains} showcases that the estimation performance with both the FC and PC cases with $N_{\rm RF}=2$ worsens with respect to the $N_{\rm RF}=4$ case, a trend that was also evident in Figs.~\ref{fig:convergence_30dB} and~\ref{fig:convergence_10dB}.

To summarize, Figs.~\ref{fig:mseVsnr_4RFchains} and~\ref{fig:mseVsnr_2RFchains} verify that, reducing the number of RX RF chains at the considered semi-passive RIS leads to a decrease in channel estimation performance, suggesting that the system benefits from a more streamlined configuration while maintaining estimation accuracy. This behavior highlights the efficiency of the proposed estimation framework in adapting to different numbers of RX RF chains in the presented semi-passive RIS hardware architecture. In addition, an intriguing similarity in the NMSE performance evaluation values between the FC case with $N_{\rm RF}=1$ and the PC one with $N_{\rm RF}=2$ can be seen in Fig.~\ref{fig:mseVsnr_2RFchains}. This indicates that the additional RX RF chain in the latter case, which is the main case studied in this chapter, compensates the fact that not all RIS unit elements are attached to a RX RF chain(s). This implies that a semi-passive RIS designer can choose between less elements connections to RX RF chains and more of such chains, or FC networks with fewer number of RX RF chains, depending on requirements on the overall complexity and estimation performance.    

\begin{figure}[!t]
    \centering
    \includegraphics[scale=0.65]{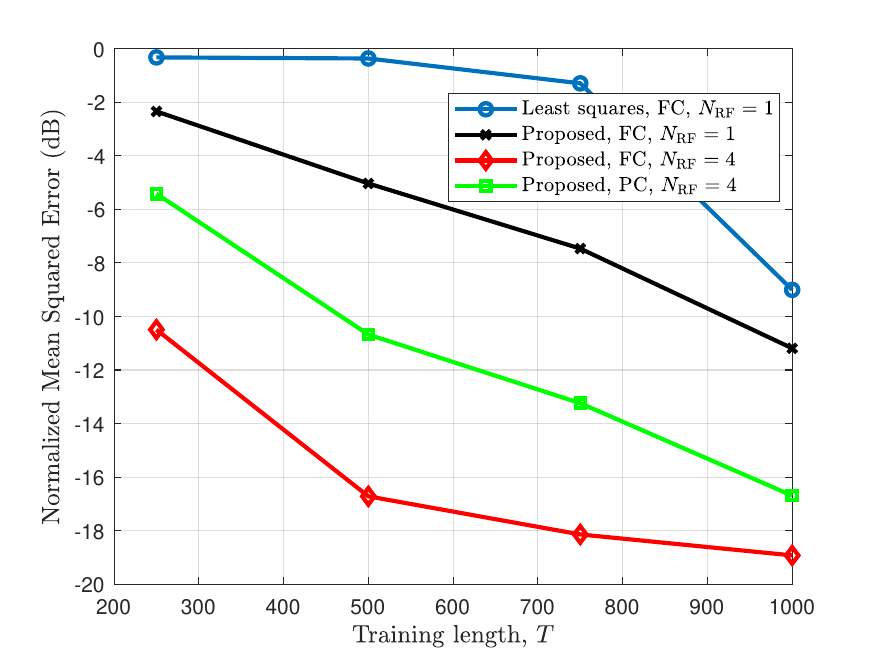}
    \caption{NMSE performance versus the training length $T$ for the parameters' setting: $N_{\rm RIS}=256$, $N_1=N_2=3$, $P=Q=3$, and the ${\rm SNR}$ value $30$ dB. Receiving RISs with $N_{\rm RF}=\{1,4\}$ RX RF chains have been considered, and the number of the absorption phase levels of each of their unit elements was set to $\vert \mathcal{W}_{\rm a} \vert = 5$.}
    \label{fig:mseVtraininglength_30}
\end{figure}

\begin{figure}[!t]
    \centering
    \includegraphics[scale=0.65]{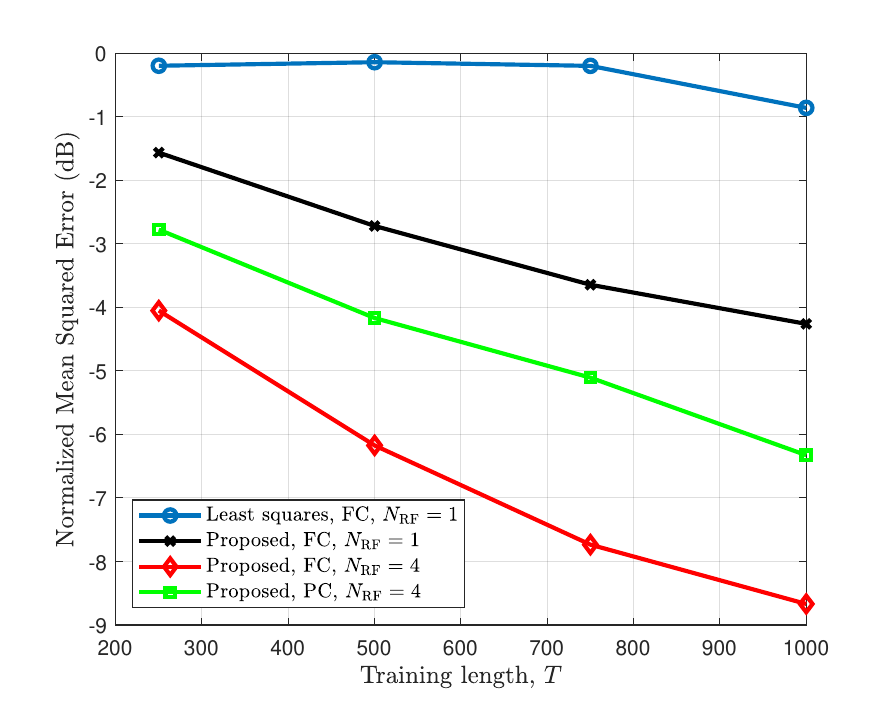}
    \caption{Same as in Fig.~\ref{fig:mseVtraininglength_30}, but for $N_{\rm RIS}=144$ and the ${\rm SNR}$ value $10$ dB.}
    \label{fig:mseVtraininglength_10}
\end{figure}

In Figs.~\ref{fig:mseVtraininglength_30} and~\ref{fig:mseVtraininglength_10}, the NMSE estimation performance as a function of the number of training/pilot symbols $T$ used in the estimation process, that took place at the ${\rm SNR}$ values of $30$~dB and $10$~dB, respectively, is illustrated. In Fig.~\ref{fig:mseVtraininglength_30}, the receiving RIS configurations of Fig.~\ref{fig:mseVsnr_4RFchains} were used, whereas, in Fig.~\ref{fig:mseVtraininglength_10}, those configurations were amended to have $N_{\rm RIS}=144$. In both figures, UE$_1$ and UE$_2$ with $N_1=N_2=3$ antenna elements and $P=Q=3$ channel propagation paths for both links were considered. It can be observed from both figures that, due to the inherently limited number of RX RF chains available at the semi-passive RIS, accurately recovering the large MIMO channel requires a significantly higher number $T$ of training symbols to achieve reliable estimation performance. This necessity arises from the fact that, fewer RX RF chains restrict the amount of spatial information captured, making the estimation process more challenging. For this reason, the least-squares approach, which is structure agnostic, exhibits the poorest estimation performance, since it does not leverage any prior knowledge about the system characteristics. Similarly, it is shown that the proposed ADMM-based channel estimation Algorithm~\ref{algorithm:proposed_algorithm_1} for the FC configuration with $N_{\rm RF}=1$ struggles to achieve satisfactory accuracy, reinforcing the notion that a single-RF receiving RIS is insufficient for effective channel recovery in extremely large MIMO system settings. Conversely, when $N_{\rm RF}=4$, the proposed approach with both FC and PC settings demonstrates a clear advantage that ameliorates with increasing $T$ values. As it can be seen, for the same $N_{\rm RF}$ and $T$ values, the FC configuration outperforms the PC one; this is a trend that was also seen in Figs.~\ref{fig:mseVsnr_4RFchains} and~\ref{fig:mseVsnr_2RFchains}. Overall, it is evident that the joint configuration of the semi-passive RIS architecture, the training length, and the operating ${\rm SNR}$ can yield different trade-offs between complexity and channel estimation precision.

\begin{figure}[!t]
    \centering
    \includegraphics[scale=0.65]{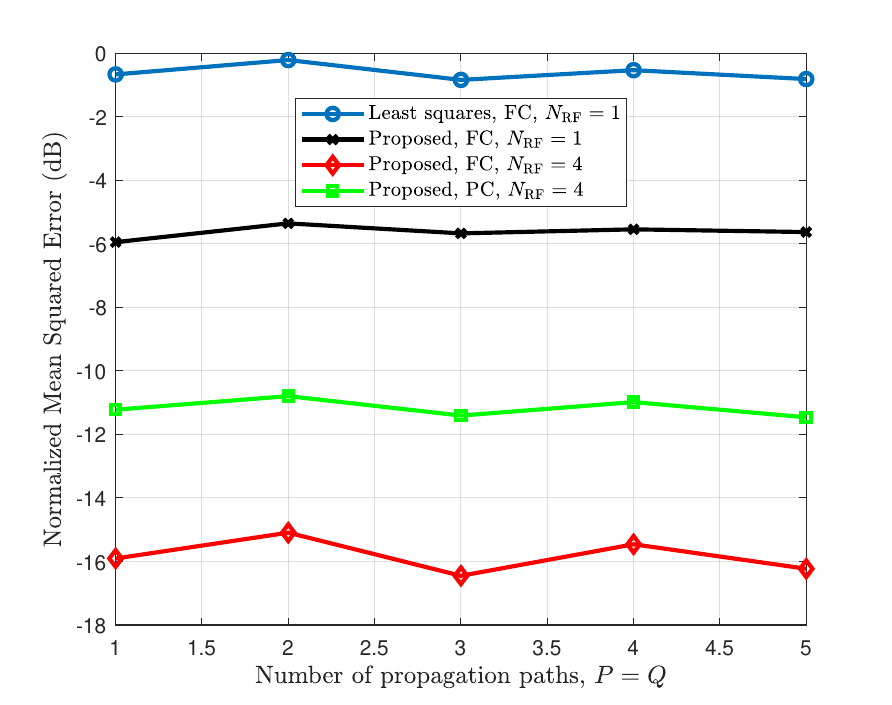}
    \caption{NMSE performance versus the number of channel propagation paths $P=Q$ for the parameters' setting: $N_{\rm RIS}=256$, $N_1=N_2=3$, $T=500$, and the ${\rm SNR}$ value $30$ dB. Receiving RISs with $N_{\rm RF}=\{1,4\}$ RX RF chains have been considered, and the number of the absorption phase levels of each of their unit elements was set to $\vert \mathcal{W}_{\rm a} \vert = 5$.}
    \label{fig:mseVmpc}
\end{figure}

Finally, Fig.~\ref{fig:mseVmpc} depicts the NMSE performance of the considered channel estimation schemes at the ${\rm SNR}$ value of $30$ dB with $T=500$ pilot symbols versus the number of channel propagation paths $P=Q$, considering the receiving RIS configurations of Fig.~\ref{fig:mseVsnr_4RFchains} and $N_1=N_2=3$ antennas elements for the UE$_1$ and UE$_2$. The results indicate that, regardless of the number of the paths, the NMSE remains stable, exhibiting consistent performance across all tested scenarios. This stability suggests that the proposed ADMM-based channel estimation approach effectively adapts to increasing path diversity without significant degradation in estimation accuracy. Thus, the proposed approach maintains robustness in diverse propagation environments. This characteristic is particularly valuable in practical applications, where multipath effects can vary significantly depending on the communication scenario. It is also shown that, increasing $N_{\rm RF}$ enhances estimation accuracy, with the FC reception architecture yielding the best performance. However, as also previously showcased, a receiving RIS relying on the PC reception architecture with reduced number of RX RF chains can still demonstrate effectiveness, particularly when balanced with an overall optimized system setup.

\subsection{Rate Results with RIS Reflection Optimization}
\begin{figure}[!t]
    \centering
    \includegraphics[scale=0.65]{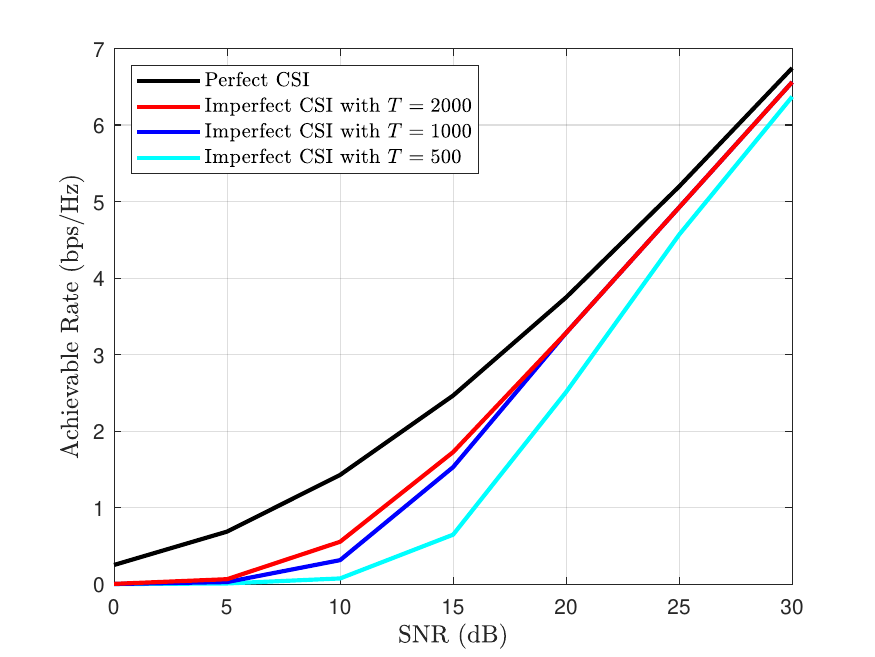}
    \caption{Achievable ergodic rate performance versus the ${\rm SNR}$ for the parameters' setting: $N_{\rm RIS}=100$, $N_1=8$, $N_2=2$, and $P=Q=3$, considering different numbers of training symbols $T$ for channel estimation with a receiving RIS having $N_{\rm RF}=1$ RX RF chain and $\vert\mathcal{W}_{\rm r}\vert=\infty$. Channel estimation took place for each SNR value indicated in the $x$-axis, and the channel estimates were then used for computing the capacity-achieving RIS reflection phase configuration via the approach presented in Section~\ref{sec:RIS_reflection_optimization}.}
    \label{fig:Rates}
\end{figure}
\begin{figure}[!t]
    \centering
    \includegraphics[scale=0.65]{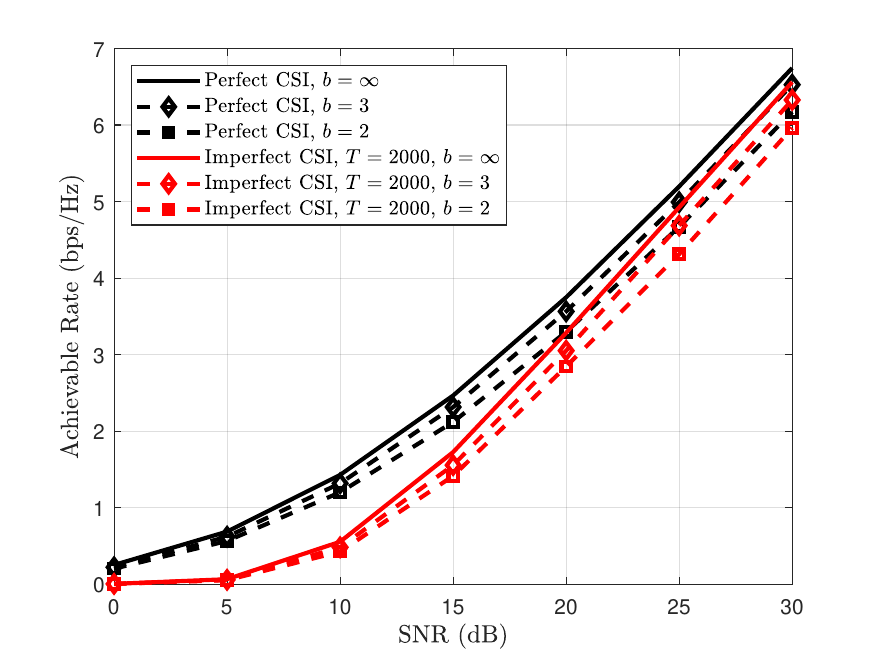}
    \caption{Same as in Fig.~\ref{fig:Rates}, but only for channel estimation with $T=2000$ and $b=\{2,3\}$ reflection quantization levels for each of the RIS unit elements (i.e., $\vert\mathcal{W}_{\rm r}\vert=\{3,5\}$). The case of continuous reflection responses per element (i.e., $b=\infty$, hence, $\vert\mathcal{W}_{\rm r}\vert=\infty$) is repeated here from Fig.~\ref{fig:Rates} for comparison purposes.}
    \label{fig:Rates_quant}
\end{figure}

We now present numerically evaluated results for the achievable rate performance of the proposed RIS-empowered MIMO communication system using the RIS reflection phase configuration scheme presented in Section~\ref{sec:RIS_reflection_optimization}. We have particularly computed the ergodic rate via the evaluation of the objective function of $\mathcal{OP}_1$ over Monte Carlo simulations, as described in the beginning of this section. In particular, using the channel estimates $\hat{\mathbf{H}}_1$ and $\hat{\mathbf{H}}_2$, which become available at the baseband processing unit of the considered semi-passive RIS hardware architecture by the end of each first phase of the proposed communication protocol (see Figs.~\ref{fig:R-RIS_system_model} and~\ref{fig:R-RIS_protocol}), an $\mathcal{OP}_1$ version with these estimates in the places of $\mathbf{H}_1$ and $\mathbf{H}_2$ is formulated, and then solved to obtain the optimized RIS reflection phase configuration. Then, this reflection configuration is inserted into the $\mathcal{OP}_1$'s objective function with the actual $\mathbf{H}_1$ and $\mathbf{H}_2$ to deduce the achievable rate with the proposed ADMM-based channel estimation approach; this is the ``Imperfect CSI'' case included in Figs.~\ref{fig:Rates} and~\ref{fig:Rates_quant} that follow. In these figures, for comparison purposes, the capacity-achieving RIS reflection phase configuration with perfect channel knowledge via Section~\ref{sec:RIS_reflection_optimization}, termed as ``Perfect CSI,'' has been also included, serving as an upper bound for the achievable rate performance.

In the conducted simulations, a semi-passive RIS with $N_{\rm RIS} = 100$ unit elements was considered, while UE$_1$ and UE$_2$ consisted of $N_1 = 8$ and $N_2 = 2$ antenna elements, respectively. To focus on the RIS-enabled link, the direct channel between UE$_1$ and UE$_2$ was assumed completely blocked, and it was thus set as $\mathbf{H} = \mathbf{0}_{N_2\times N_1}$. In addition, all three nodes were placed in an isosceles triangle, with UE$_1$ and UE$_2$ forming its basis, while the RIS was placed on the triangle's upper corner, setting the two other sides of the triangle equal to $d = d_{\rm UE_1,RIS} = d_{\rm RIS,UE_2} = 5\,m$. The pathloss was set as $d^{-a}$ with $a = 2.2$ being the pathloss exponent. The step size of Algorithm~\ref{algorithm:RIS_optimization} was chosen as $\mu=2.5$, and we considered both continuous (i.e., $[\bar{\boldsymbol{\varphi}}]_i\in\mathcal{W}_{\rm r,c}$ $\forall$$i=1,2,\dots,N_{\rm RIS}$, thus, $b = \infty$) and finite discrete (i.e., $[\bar{\boldsymbol{\varphi}}]_i\in\mathcal{W}_{\rm r}$ with $b=\{2,3\}$ $\forall$$i=1,2,\dots,N_{\rm RIS}$) reflection states per RIS unit element, as well as the training lengths $T \in \{500,1000,2000\}$ for the proposed ADMM-based channel estimation approach. For all results that follow, we have generated $200$ independent channel realizations for $\mathbf{H}_1$ and $\mathbf{H}_2$, and then the channel estimates $\hat{\mathbf{H}}_1$ and $\hat{\mathbf{H}}_2$ were obtained for different values of the ${\rm SNR}$ and $T$.
\begin{figure}[!t]
    \centering
    \includegraphics[scale=0.65]{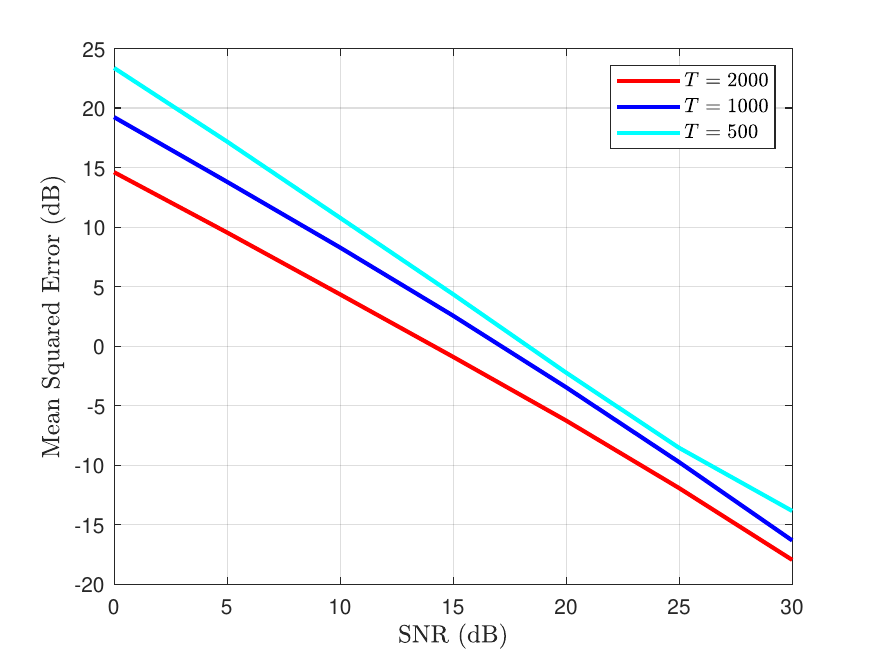}
    \caption{MSE performance of the channel estimation considered in Fig.~\ref{fig:Rates} for the different $T$ values.}
    \label{fig:Errors}
\end{figure}

Figure~\ref{fig:Rates} illustrates the achievable ergodic rate performance versus the ${\rm SNR}$ for different numbers of pilot symbols $T$ used with the proposed ADMM-based channel estimation approach, considering a semi-passive RIS equipped with $N_{\rm RF}=1$ RX RF chain and the set $\mathcal{W}_{\rm r,c}$ of continuous reflection phase states per unit element. As expected, the rate follows a non-decreasing trend as the SNR increases for both the perfect and imperfect CSI cases. More importantly, it is evident that, as $T$ increases, the gap between the ideal case of perfect CSI and that with the estimated one becomes smaller, especially in the high SNR regime, i.e., for ${\rm SNR} > 20$ dB. In those cases, a large $T$ value and a large ${\rm SNR}$ facilitate channel estimation. This is evident in Figs.~\ref{fig:mseVtraininglength_30} and \ref{fig:mseVtraininglength_10}, where it was shown that, for larger number $T$ of pilot symbols, the estimation process improves in terms of NMSE performance. In addition, Figs.~\ref{fig:mseVsnr_4RFchains} and \ref{fig:mseVsnr_2RFchains} demonstrate that the NMSE admits smaller values as the ${\rm SNR}$ increases. As a result, the combination of the latter two reasons explains the observed gap in the low ${\rm SNR}$ range, which is also depicted in Fig.~\ref{fig:Errors} for the simulation setup considered in Fig.~\ref{fig:Rates}. 

On the other hand, the impact of the number of quantization bits $b$ for the RIS reflection response per unit element on the achievable rate is illustrated in Fig.~\ref{fig:Rates_quant}, where the cases of $b=2$ and $b=3$ for both the scenarios of perfect CSI and imperfect CSI with $T=2000$ were considered. Evidently, it can be observed that, for increasing $b$, the performance loss with respect to the achievable rate when $b=\infty$ decreases. However, even for $b=2$, the rate loss due to quantization at the elements of the reflection phase configuration vector  $\bar{\boldsymbol{\varphi}}$ is negligible in the high SNR regime (i.e. for ${\rm SNR} \geq 20$ dB). This trend witnesses that, for moderate to high ${\rm SNR}$ values and wireless channels with small numbers of propagation paths, both the presented channel estimation and the reflection phase configuration schemes, which have been designed for the proposed single-RF receiving RIS hardware architecture, perform sufficiently well.  

\section{Conclusions}\label{sec:conclusions}
In this chapter, we considered a semi-passive RIS hardware architecture incorporating RX RF chains and a baseband processing unit, and presented an approach to utilize it for efficiently optimizing the performance of RIS-assisted MIMO communication systems. In particular, we presented a channel estimation protocol according to which the RIS receives non-orthogonal training pilot sequences by two multi-antenna UEs via tunable absorption phase profiles, and then, estimates the respective channels via its signal processing unit. The channel estimates were then used by the RIS controller to design its capacity-achieving reflection phase configuration. The proposed ADMM-based channel estimation algorithm profits from the RIS random spatial absorption sampling to capture the entire signal space, and exploits the beamspace sparsity and low-rank properties of large MIMO channels, which is particularly relevant for high-frequency communication systems. Our extensive numerical investigations showcased the superiority of the proposed channel estimation technique over benchmark schemes for various system and RIS hardware configuration parameters, as well as the effectiveness of using channel estimates at the RIS side to dynamically optimize the reflection coefficients of its unit elements.

A multitude of interesting research directions arises for receiving RISs capable of estimating parameters of their impinging waveforms, following also the contributions of this chapter. For example, the joint realization of tunable reflection and absorption phase responses is still an open research problem that heavily depends on the targeted frequency band as well as the operation bandwidth~\cite{GGM_ACCESS_2024}. Another research direction deals with the extension of the current channel estimation framework to more than two UEs, also considering physics-/circuit-compliant models for the semi-passive RIS and the resulting RIS-parameterized channels. To this end, it is interesting to extend the proposed communication protocol to incorporate contention mechanisms as well as schemes capable of detecting the presence of an available for communications semi-passive RIS in the smart wireless environment~\cite{RIS_detection2025}. Last but not least, the implications, in terms of pilot transmission overhead and computational complexity, of non-explicit estimation of the involved channel matrices need to be identified (e.g., principal channel components~\cite{MAD_TWC_2023,MA2024} or directions of strong channel paths~\cite{locrxris_all}) to unveil the degrees of freedom in designing receiving RISs with minimal number of RX RF chains ~\cite{AV_ICASSP_2020}, of possibly reduced complexity hardware components~\cite{GAA2024}. Incorporating realistic losses in metasurface-based reception systems, like receiving RISs, is another critical research direction~\cite{DMA_Losses_2025}.

\section*{Acknowledgements}
This work was supported by the Smart Networks and Services Joint Undertaking (SNS JU) projects TERRAMETA and 6G-DISAC under the European Union's Horizon Europe research and innovation programme under Grant Agreement numbers 101097101 and 101139130, respectively. TERRAMETA also includes top-up funding by UK Research and Innovation (UKRI) under the UK government's Horizon Europe funding guarantee. 

\bibliographystyle{splncs03_unsrt}
\bibliography{references}
\end{document}